\documentclass[a4paper,fleqn]{cas-dc}

\usepackage[numbers]{natbib}

\usepackage{amsmath}	
\def\tsc#1{\csdef{#1}{\textsc{\lowercase{#1}}\xspace}}
\tsc{WGM}
\tsc{QE}
\tsc{EP}
\tsc{PMS}
\tsc{BEC}
\tsc{DE}
\def\physrep{Physics Reports}
\def\aap{Astronomy \& Astrophysics}
\def\aj{Astronomical Journal}

\def\mnras{MNRAS}
\def\jcap{JCAP}
\def\prd{Physical Review D}
\def\prl{Physical Review Letters}
\def\apj{Astrophysical Journal}

\def\na{New Astronomy}

\begin{document}
\let\WriteBookmarks\relax
\def\floatpagepagefraction{1}
\def\textpagefraction{.001}
\shorttitle{PAge-like Unified Dark Fluid model}
\shortauthors{Y. Su et~al.}

\title [mode = title]{Physical Implications and Updated Observational Constraints of the PAge-like Unified Dark Fluid Model}                      

\tnotetext[1]{
   This work is supported by the National Natural Science Foundation of China (NSFC) under its Key Program (Grant No. 12533002) and General Program (Grant No. 12073088), the National key R\&D Program of China (Grant No. 2020YFC2201600), the National SKA Program of China No. 2020SKA0110402, and Guangdong Basic and Applied Basic Research Foundation (Grant No.2024A1515012573).}


\author[1,2]{Su Yan}[style=chinese,
                        orcid=0009-0003-1993-0769
                        ]
\affiliation[1]{organization={School of Physics and Astronomy, Sun Yat-sen University},
                addressline={2 Daxue Road}, 
                city={Zhuhai},
                postcode={519082}, 
                state={Guangdong},
                country={China}}
\affiliation[2]{
organization={CSST Science Center for the Guangdong-Hongkong-Macau Greater Bay Area, Sun Yat-sen University},
                addressline={}, 
                postcode={519082}, 
                city={Zhuhai},
                country={China}}

\author[1,2]{Huang Zhiqi}[style=chinese]
\cormark[1]
\ead{huangzhq25@mail.sysu.edu.cn}
\cortext[cor1]{Corresponding author}

\author[1,2]{Yao Yanhong}[style=chinese]
\author[1,2]{Wang Junchao}[style=chinese]
\author[1,2]{Liu Jianqi}[style=chinese]




\begin{abstract}
The standard paradigm of cosmology assumes two distinct dark components, namely dark matter and dark energy. However, the necessity of splitting the dark-side world into two sectors has not been experimentally or theoretically proven. Unified dark fluid models provide an alternative in which a single fluid accounts for both phenomena. It is shown in Wang et al. 2024 that a PAge-like unified dark fluid (PUDF) can explain both the cosmic microwave background (CMB) and late-universe data, with the fitting quality not much worse than the standard Lambda cold dark matter ($\Lambda$CDM) model. 
Using the Planck 2018 CMB, baryon acoustic oscillations measurement from the dark energy spectroscopic instrument (DESI) data release 2, dark energy survey 5-year supernova data, and cosmic-chronometer data, we update the constraints on PUDF and clarify its physical implications.
We show that PUDF can reproduce the primary CMB anisotropies, the background expansion history, and linear growth that are very close to the $\Lambda$CDM prediction. Nevertheless, the combined datasets still favor $\Lambda$CDM, largely due to the significant tension between CMB and DESI + SNe data, which exceeds the $4\sigma$ level in PUDF and remains non-negligible in the $w$CDM framework. Using mock data generated from the Planck best-fit $\Lambda$CDM model, we find that PUDF and $\Lambda$CDM cannot be statistically distinguished, indicating that the precision of current data is insufficient to separate the two models. Overall, the apparent preference for $\Lambda$CDM may be driven by dataset inconsistencies rather than a genuine physical difference, leaving unified dark fluid models as viable alternatives within current observational limits. 

\end{abstract}


\begin{highlights}






\item Unified dark fluid models remain viable alternatives to $\Lambda$CDM under current data
\item Current observations cannot statistically distinguish PUDF from $\Lambda$CDM
\item The apparent $\Lambda$CDM preference is driven by CMB–DESI+SNe dataset tension
\item Dataset inconsistencies limit physical discrimination between dark sector models
\item Improved data consistency is essential to test unified dark sector scenarios
\end{highlights}


\maketitle

\section{Introduction}\label{sec:intro}
Our universe contains approximately 5\% baryonic matter and 95\% dark components which are commonly considered as dark matter and dark energy~\citep{aghanim2020planck}. Dark matter plays an important role in the formation of large scale structures, while dark energy drives the accelerated expansion of the universe~\citep{1998AJ....116.1009R, 1999ApJ...517..565P}. In the standard Lambda cold dark matter ($\Lambda$CDM) model, dark energy is interpreted as the cosmological constant ($\Lambda$) or equivalently the vacuum energy. The cosmological constant interpretation of dark energy has a fine-tuning problem, which questions the smallness of vacuum energy density~\citep{Weinberg1989}, and a coincidence problem, which asks why the vacuum energy density is the same order of magnitude as the matter density today~\citep{Zlatev1999}. The fine-tuning and coincidence problems also apply to many alternative models of dark energy~\citep{Jerome2012, Joyce2015}. 

The coincidence between the densities of dark matter and baryonic matter is usually considered to be less problematic, as baryons and dark matter may have a similar origin in the early universe. Thus, the coincidence problem of dark energy could be naturally resolved if we unify dark energy and dark matter into one single component that shares a common origin with baryonic matter. To explain the cosmological data, the unified dark component should behave like pressure-less dust in the early (redshift $z\gg 1$) universe and should have negative pressure in the late ($z\lesssim 1$) universe. If the dust-to-$\Lambda$ transition could be triggered by the inhomogeneity of the unified dark component itself, or by its coupling to neutrinos which becomes non-relativistic in the late-universe, the fine-tuning problem would also be resolved.

Beyond enduring conceptual difficulties, the vacuum-energy paradigm for dark energy has been further called into question by recent advances in observational cosmology. Recent multi-probe analyses incorporating the Dark Energy Spectroscopic Instrument (DESI) baryon acoustic oscillations (BAO) measurements, Type Ia supernovae (SNe) luminosity distances, and cosmic microwave background (CMB) power spectra provide strong empirical support for phantom-crossing dark energy scenarios~\cite{DESI:2024mwx,DESI:2025zgx}. Theoretically, constructing such phantom-divide crossing behavior within a single-component framework poses significant challenges, as canonical scalar field implementations typically develop quantum instabilities when approaching the phantom threshold (pressure-to-density ratio less than $-1$) due to effective negative kinetic energy terms. Notably, the effective equation of state (total pressure divided by total density) for the combined dark sector (dark matter plus dark energy) in the DESI BAO+SNe+CMB best-fit model remains above $-1$ across cosmic history. This again motivates the unification paradigm wherein dark matter and dark energy emerge as different manifestations of a unified dark component, thereby naturally circumventing the phantom-crossing problem. For concrete model-building in this direction, see also~\citet{Kou:2025yfr}.

While it is difficult to formulate a fundamental theory to implement all the aforementioned ideas, it is possible to construct an effective action or to build a phenomenological model with fluid approximation. Examples of unified dark-sector scenarios include Chaplygin gas and its many  variations~\citep{kamenshchik2001alternative,bento2002generalized,bilic2002unification,zhang2006new,xu2012revisiting,li2013viscous,Xu2014,Kumar2014,lu2015,Ferreira2018,abdullah2022growth,mandal2024dynamical,dunsby2024unifying, Hashim2025preprint, Fortunato2025}, scalar field with non-canonical kinetic energy~\citep{scherrer2004purely,guendelman2016unified,Sahni2017,bertacca2008scalar,bertacca2011unified,mishra2021unifying,chavanis2022k,frion2024bayesian}, modified gravity theories~\citep{liddle2006inflation,henriques2009unification,Tripathy2015,koutsoumbas2018unification,dutta2018cosmological,Tripathy2020,sa2020unified,Gadbail2022,Shukla2025},  quark bag model~\citep{Brilenkov2013}, Bose-Einstein condensate~\citep{Das2023}, polytropic dark matter~\citep{Kleidis2015}, and other fluid models~\citep{colistete2007bulk,dou2011bulk,elkhateeb2019dissipative,elkhateeb2023dissipative, Wang:2024rus}. Although some of the models have difficulties to predict cosmological perturbations that fit the current data~\citep{Sandvik2004,Gorini2008,Radicella2014,Cuzinatto2018,Quiros2025preprint}, it has been shown numerically that a unified dark fluid with negligible anisotropic stress and zero sound speed in general can make $\Lambda$CDM-like predictions at the background and linear-perturbation levels~\citep{Davari2018,Wang:2024rus}. 

In the PAge-like unified dark fluid (PUDF) model that was proposed in \citet{Wang:2024rus}, the unified dark component is assumed to be a fluid with a smooth background evolution parameterized by the PAge approximation~\citep{Huang:2020mub,Huang-2108-03959}. The PAge approximation is based on two assumptions, that the dark component(s) behave like dust at high-redshift, and that the dimensionless combination $Ht$, where $H$ is the Hubble parameter and $t$ is the age of the universe, is a slowly varying smooth function of $t$. The minimal PUDF contains seven cosmological parameters, with the standard $\Omega_ch^2$ (CDM density) replaced by the PAge parameters $p_{\rm age}$ ($\sim$ age of the universe) and $\eta$ (deviation from Einstein-de Sitter universe). By modifying the Boltzmann code CLASS~\citep{Diego_Blas_2011}, \citet{Wang:2024rus} computed the linear perturbations in PUDF and found that PUDF can give predictions similar to those of $\Lambda$CDM. Further analysis of Bayesian evidence shows that $\Lambda$CDM is favored over PUDF by the current cosmological data including CMB, BAO, SNe, and cosmic chronometers (CC)~\citep{Wang:2024rus}.

The results found in \citet{Wang:2024rus}, however, lack a clear physical interpretation. It is unclear to what extent PUDF can mimic $\Lambda$CDM at the background and linear-perturbation levels. Neither do we know what key difference between PUDF and $\Lambda$CDM has led to the slightly different $\chi^2$ fits to the data. Similar problems exist for the earlier work \citet{Davari2018} with a polynomial-based parameterization. Therefore, this work aims to improve the theoretical understanding of the similarities and nuances between PUDF and $\Lambda$CDM, and updates the results with the latest datasets.
While the theoretical exploration is done in Section~\ref{sec:model}, we revisit the Bayesian parameter inference and update the results in Section~\ref{sec:data}.
Section~\ref{sec:concludions} summarizes and concludes.

Throughout the paper we work with the spatially flat background metric $\mathrm{d}s^2 =\mathrm{d}t^2 - a(t)^2\mathrm{d}\mathbf{x}^2$, where the scale factor $a(t)$ is related to the cosmological redshift $z$ via $a=\frac{1}{1+z}$. The Hubble parameter is defined as $H(t) = \frac{\dot a}{a}$, where a dot denotes derivative with respect to the background time $t$. We use a subscript $0$ to denote quantities at redshift zero. For example, the Hubble constant $H_0$ is the Hubble parameter at redshift zero, often written as $100h\,\mathrm{km\cdot s^{-1}Mpc^{-1}}$. The critical density is defined as $\rho_{\rm crit} = \frac{3H_0^2}{8\pi G}$, where $G$ is Newton's gravitational constant. We use subscripts $b$, $c$, $d$, $\nu$, $\gamma$, $\Lambda$ for baryon, cold dark matter, unified dark fluid, neutrinos, photons and vacuum energy, respectively. For a component $X = b, c, d, \nu,\gamma,\Lambda$, the abundance parameter $\Omega_X$ is defined as the ratio between its current background density $\rho_{X0}$ and the critical density $\rho_{\rm crit}$. For parameter inference, unless otherwise specified, we assume flat priors on the logarithm amplitude of primordial scalar perturbations $\ln(10^{10}A_s)$, the tilt of primordial scalar perturbations $n_s$, the reionization optical depth $\tau_{\rm re}$, the angular extension of the sound horizon at recombination $\theta_*$, the baryon density $\Omega_b h^2$, and the parameter(s) for the dark component(s), i.e., $\Omega_ch^2$ for $\Lambda$CDM and $(p_{\rm age}, \eta)$ for PUDF. For the neutrino masses, we assume a massive species with minimum mass $0.06\,\mathrm{eV}$ and two massless species. In the context of the $\Lambda$CDM model, we define the matter abundance $\Omega_m = \Omega_b+\Omega_c$ for brevity. Here we do not include $\Omega_\nu$ in the definition of $\Omega_m$ because we are more interested in matching matter density at high redshift where neutrinos are relativistic. 

\section{Theoretical Comparison between PUDF and $\Lambda$CDM \label{sec:model}}

\subsection{{\rm PUDF} basics}

PUDF generalizes the original PAge approximation by adding the radiation and neutrino contribution at high redshift. The Hubble parameter is given by
\begin{eqnarray}
H^2(z) &=& H_0^2\left[\Omega_\gamma + \sum_{i=1}^3 \Omega_{\nu,i}\frac{\mathrm{I}_{\rho}\left(\frac{m_{\nu,i} }{(1+z)T_{\nu}}\right)}{\mathrm{I}_{\rho}\left(\frac{m_{\nu,i}}{T_{\nu}}\right)}\right](1+z)^4 \nonumber \\
&& +H_{\rm PAge}^2(z), \label{eq:H}     
\end{eqnarray}
where $m_{\nu, i}$ is the neutrino mass of the $i$-th species; $T_{\nu}=T_{\rm CMB}\left(\frac{4}{11}\right)^{1/3}\approx 1.95\,\mathrm{K}$ is the effective temperature for neutrino momentum distribution. The neutrino density integral is
\begin{equation}
  \mathrm{I}_\rho(\lambda) \equiv  \frac{1}{2\pi^2}\int_0^\infty \frac{x^2\sqrt{x^2+\lambda^2}}{e^x+1}~\mathrm{d}x.
\end{equation}
The contribution from baryon and dark fluid is encoded in the $H_{\rm PAge}^2(z)$ term. The function $H_{\rm PAge}(z)$ is given by two parameters $(p_{\rm age}, \eta)$ and an auxiliary variable $\beta$ running from $0$ to $p_{\rm age}$.
\begin{eqnarray}
H_{\mathrm{PAge}} &=& H_0\sqrt{1-\Omega_\nu - \Omega_\gamma} \nonumber \\
&& \times \left[1+\frac{2}{3}\left(1-\eta\frac{\beta}{p_\mathrm{age}} \right)\left(\frac{1}{\beta}-\frac{1}{p_\mathrm{age}} \right)\right]. \label{eq:HPAge}  \\
 1+z &=& \left(\frac{p_{\rm age}}{\beta}\right)^{2/3}  \times \exp \bigg\{-\frac{\eta}{3} \left[\left(\frac{\beta}{p_{\rm age}}\right)^2 - 1\right] \nonumber \\
 && - \left[p_{\rm age}-\frac{2}{3}(1+\eta)\right]\left(\frac{\beta}{p_{\rm age}}-1\right)\bigg\}. \label{eq:z}
\end{eqnarray}
Here the parameter $p_{\rm age}$ is approximately the age of the universe in unit of $H_0^{-1}$ and $\eta$ is a phenomenological parameter describing the deviation from the Einstein-de Sitter universe. The running variable $\beta$ is approximately $H_0t$. When radiation and neutrinos can be neglected and the relation $\beta = H_0 t$ holds exactly, Eq.~\eqref{eq:HPAge} reduces to the original PAge formalism introduced in Ref.~\cite{Huang:2020mub}. Equation~\eqref{eq:z} follows from integrating Eq.~\eqref{eq:HPAge} with respect to $\beta = H_0 t$ and then exponentiating both sides. In PUDF, the presence of radiation and neutrinos causes $H_{\rm PAge}$ to deviate from the actual Hubble parameter and also makes $\beta$ slightly differ from $H_0 t$. 

It should be emphasized that Eq.~\eqref{eq:z} is not an exact reformulation of the $\Lambda$CDM Friedmann equation, but a phenomenological parameterization of the expansion history. 
Nevertheless, the original PAge studies have shown that this parameterization can accurately approximate the expansion histories of a wide range of cosmological models, including the $\Lambda$CDM model~\cite{Huang:2020mub,luo2020reaffirming}.

The density of the unified dark fluid is given by
\begin{equation}
\rho_d(z)=\frac{3}{8\pi G}H_\mathrm{PAge}^2-\rho_b(z),
\end{equation}
where the physical baryon density $\rho_b(z)$ is
\begin{equation}
    \rho_b(z) = \rho_{\rm crit}\Omega_b(1+z)^3\propto \Omega_bh^2(1+z)^3.
\end{equation}
The pressure of the dark fluid, $p_d$, is derived from the continuity equation
\begin{equation}
    \dot{\rho}_d+3H(\rho_d+p_d)=0,
\end{equation}
and the equation of state for the unified dark fluid is given by
\begin{equation}
    w \equiv \frac{p_d}{\rho_d} = \frac{1+z}{3\rho_d}\frac{\mathrm{d}\rho_d}{\mathrm{d}z}-1.  \label{eq:w}
\end{equation}
While Equations~(\ref{eq:HPAge}-\ref{eq:w}) may appear rather complicated, they stem from just two straightforward assumptions~\citep{Huang:2020mub}: (1) the dark component behaves like dust at high redshifts $z\gg 1$, and (2) the dimensionless quantity $Ht$ can be approximated by a quadratic function of $t$.  The Taylor expansion of $Ht$ is conceptually similar to expanding the dark energy equation of state $w(a)$, but the PUDF framework offers two key advantages over the $w_0w_a$CDM model: it is more economical, requiring one fewer parameter, and better physically motivated, as $Ht$ has been shown to vary smoothly and slowly across a wide range of well-studied cosmological models~\citep{Huang:2020mub}.

To ensure theoretical consistency at the perturbative level, the unified dark component is formulated within a covariant framework. A macroscopic dark fluid with rest-frame energy density $\rho_d$ and rest-frame pressure $p_d$ admits a covariant description through the energy-momentum tensor
\begin{equation}
T_{\mu\nu} = (\rho_d + p_d) u_\mu u_\nu + p_d g_{\mu\nu} + \pi_{\mu\nu},
\end{equation}
where $\pi_{\mu\nu}$ is the spatial ($\pi_{\mu\nu}u^\mu = 0$) and traceless ($\pi^\mu_{\ \mu}  = 0$) viscous-stress tensor.

Such a structure can be derived from an underlying action principle. For example, in scalar-field realizations such as purely kinetic k-essence models~\citep{armendariz2001essentials,scherrer2004purely} with the action
\begin{equation}
S = \int d^4x \sqrt{-g}\, P(X),
\end{equation}
where $X = -\frac{1}{2}\nabla_\mu\phi\nabla^\mu\phi$, variation of the action with respect to the metric yields
\begin{equation}
T_{\mu\nu} = P_{,X}\nabla_\mu\phi\nabla_\nu\phi + P(X) g_{\mu\nu},
\end{equation}
which corresponds to the fluid quantities
\begin{equation}
\rho_d = 2X P_{,X} - P, \quad p_d = P.
\end{equation}
This demonstrates explicitly that the fluid description employed here can arise from a fully covariant Lagrangian formulation.

We emphasize, however, that the phenomenological parameters of the PUDF model (such as $p_{\rm age}$ and $\eta$) are not fundamental constants appearing directly in the underlying Lagrangian. Instead, they parametrize the background cosmological solutions $\rho_d(z)$ and $p_d(z)$. In other words, they characterize a particular class of solutions of a covariant fluid system rather than specifying the microscopic theory itself.

Given the covariant energy-momentum tensor, the perturbation equations follow directly from the conservation law $\nabla_\mu T^{\mu\nu} = 0$, whose linear-order perturbations in the synchronous gauge are
\begin{eqnarray}
\delta^\prime &=& -(1 + w)\left(\theta + \frac{{h^i_{\ i}}^\prime}{2}\right) - 3\frac{a^\prime}{a}\left(c_{\mathrm{s,eff}}^2 - w\right)\delta \nonumber\label{eq:delta} \\
&& - 9\left(\frac{a^\prime}{a}\right)^2\left(c_{\mathrm{s,eff}}^2 - c_{\mathrm{s,ad}}^2\right)(1 + w)\frac{\theta}{k^2},  \\
\theta^\prime &=& -\frac{a^\prime}{a}\left(1 - 3c_{\mathrm{s,eff}}^2\right)\theta + \frac{c_{\mathrm{s,eff}}^2}{1 + w}k^2\delta - k^2\sigma,\label{eq:theta} 
\end{eqnarray}
where the prime denotes the derivative with respect to conformal time $\tau\equiv\int\frac{\mathrm{d}t}{a}$,  $\delta = \delta\rho_d/\rho_d$ is the relative density perturbation, $\theta$ is the velocity divergence of the dark fluid, $k$ is the comoving wavenumber, $h^i_{\ i}$ is the trace of the metric perturbations, and $\sigma$ is the shear perturbations of the fluid which is assumed to be negligible in this work. The adiabatic sound speed of the fluid $c_\mathrm{s,ad}$ is specified as
\begin{equation}
c_\mathrm{s,ad}^2=\frac{\dot{P}}{\dot{\rho}}=w-\frac{\dot{w}}{3H\left(1+w\right)}.
\end{equation}
The most relevant quantity describing the propagation of pressure perturbations is the effective sound speed in the fluid rest frame, $c_\mathrm{s, eff}^2$, which we assume to be negligible, too. The difference between $c_\mathrm{s, eff}$ and $c_\mathrm{s, ad}$ is due to relativistic correction to the density perturbations. A component with a negative equation of state and vanishing $c_\mathrm{s, eff}$ can be realized, for example, in purely kinetic k-essence models~\citep{armendariz2001essentials, scherrer2004purely}.

Equations~(\ref{eq:delta}) and~(\ref{eq:theta}) do not form a closed system by themselves, since the trace of the metric perturbation $h^i_{\ i}$ is determined by the perturbed Einstein equations. In the synchronous gauge, the scalar metric perturbations are conventionally decomposed into a trace part $h^i_{\; i}$ and a traceless part characterized by the potential $\mu$ (usually denoted $\eta$ in the literature, but here replaced by $\mu$ to avoid conflict with the background parameter $\eta$). In Fourier space, the synchronous-gauge metric perturbations are given by
$h_{ij} = \frac{1}{3} (h^{m}_{\; m}) \,\delta_{ij} + 6\mu \left( \hat{k}_i \hat{k}_j - \frac{1}{3}\delta_{ij} \right)$, where $\hat{k}_i \equiv k_i/|\mathbf{k}|$ are the components of the normalized wave-number vector. A closed system is obtained with two additional Einstein equations
\begin{align}
k^2\mu-\frac{1}{2}\frac{a^\prime}{a}(h^i_{\ i})^\prime
&=4\pi G a^2 \delta T^0_{\ 0,\mathrm{tot}},\\
k^2\mu^\prime
&=4\pi G a^2 \sum_A(\bar\rho_A+\bar{p}_A)\theta_A ,
\end{align}
where the subscript ``tot'' denotes the sum over all species.
In our numerical implementation, the dark-fluid conservation equations are solved together with the linearized Einstein equations in the modified CLASS code~\cite{Wang:2024rus}.

\subsection{Matching the primary CMB~\label{sec:matchCMB}}

In the high-redshift limit ($z \gg 1$, $\beta \ll p_{\rm age}$), the $1/\beta$ term in the PAge parameterization strongly dominates. Expanding the exact background equations~(\ref{eq:HPAge}-\ref{eq:z}) to the linear order of $\beta$, we obtain
\begin{eqnarray}
H_{\rm PAge}^2 &\approx & \frac{4H_0^2(1-\Omega_\nu-\Omega_\gamma)}{9p_{\rm age}^2}e^{2+\eta-3p_{\rm age}} (1+z)^3 \nonumber \\
&&\times \left[1+\left(6-\frac{4(1+\eta)}{p_{\rm age}}\right)\beta\right].\label{eq:HPAge_early}
\end{eqnarray} 

In the pre-recombination epoch where $z\gtrsim 1000$, the $O(\beta)$ correction is below $10^{-4}$ level. Thus, to a very good approximation, $H_{\rm PAge}^2$ is proportional to $(1+z)^3$ and the unified dark fluid behaves like a CDM component.

If we define an effective CDM abundance
\begin{equation}
   \Omega_{c, \rm eff} =\frac{4(1-\Omega_\nu-\Omega_\gamma)}{9p_{\rm age}^2}e^{2+\eta-3p_{\rm age}} - \Omega_b, \label{eq:Omegaceff} 
\end{equation}
the physical density of the dark fluid in the pre-recombination epoch can be written in a familiar way
\begin{equation}
\rho_d \rvert_{\text{high }z}\approx \rho_{\rm crit}\Omega_{c,\rm eff} (1+z)^3 .
\end{equation}
The primary CMB power spectrum relies on the primordial seeds, the pre-recombination physics, the conversion from the physical scale on the last-scattering surface to the observed angular scale, and the scattering between CMB photons and the reionized electrons in the late universe. The parameters controlling these effects are listed in Table~\ref{tab:primaryCMB}. It is clear that if we match $\Omega_{c,\rm eff}h^2$ in PUDF to $\Omega_ch^2$ in $\Lambda$CDM,  and fix all the other parameters, PUDF and $\Lambda$CDM should predict almost identical primary CMB power spectra with a relative difference less than $O(10^{-4})$. In other words, to match the primary CMB power spectrum to $\Lambda$CDM prediction, $p_{\rm age}$ and $\eta$ should satisfy the constraint
\begin{equation}
\left.\frac{4(1-\Omega_\nu-\Omega_\gamma)}{9p_{\rm age}^2}e^{2+\eta-3p_{\rm age}} \right\vert_{\rm PUDF}= \left.\Omega_m\right\vert_{\rm \Lambda CDM}, \label{eq:CMBmatchexact}
\end{equation}
which simplifies to
\begin{equation}
\left.\frac{4}{9p_{\rm age}^2}e^{2+\eta-3p_{\rm age}} \right\vert_{\rm PUDF}= \left.\Omega_m\right\vert_{\rm \Lambda CDM}, \label{eq:CMBmatch}
\end{equation}
if $\Omega_\nu$ and $\Omega_\gamma$ are negligible.

\begin{table*}
\centering
\caption{Parameters controlling primary CMB power spectrum\label{tab:primaryCMB}}
\begin{tabular}{ll}
\hline\hline
physical effects & parameters  \\
\midrule
primordial seeds & $A_s$ and $n_s$ \\
pre-recombination physics & $\Omega_b h^2$, $T_{\rm CMB}$, neutrino masses, $\Omega_{c,\rm eff} h^2$ for PUDF or $\Omega_ch^2$ for $\Lambda$CDM \\
angular scale conversion & $\theta_*$ \\
reionization & $\tau_{\rm re}$ \\
\hline
\end{tabular}
\end{table*}

\begin{figure*}
\centering
\includegraphics[width=\textwidth]{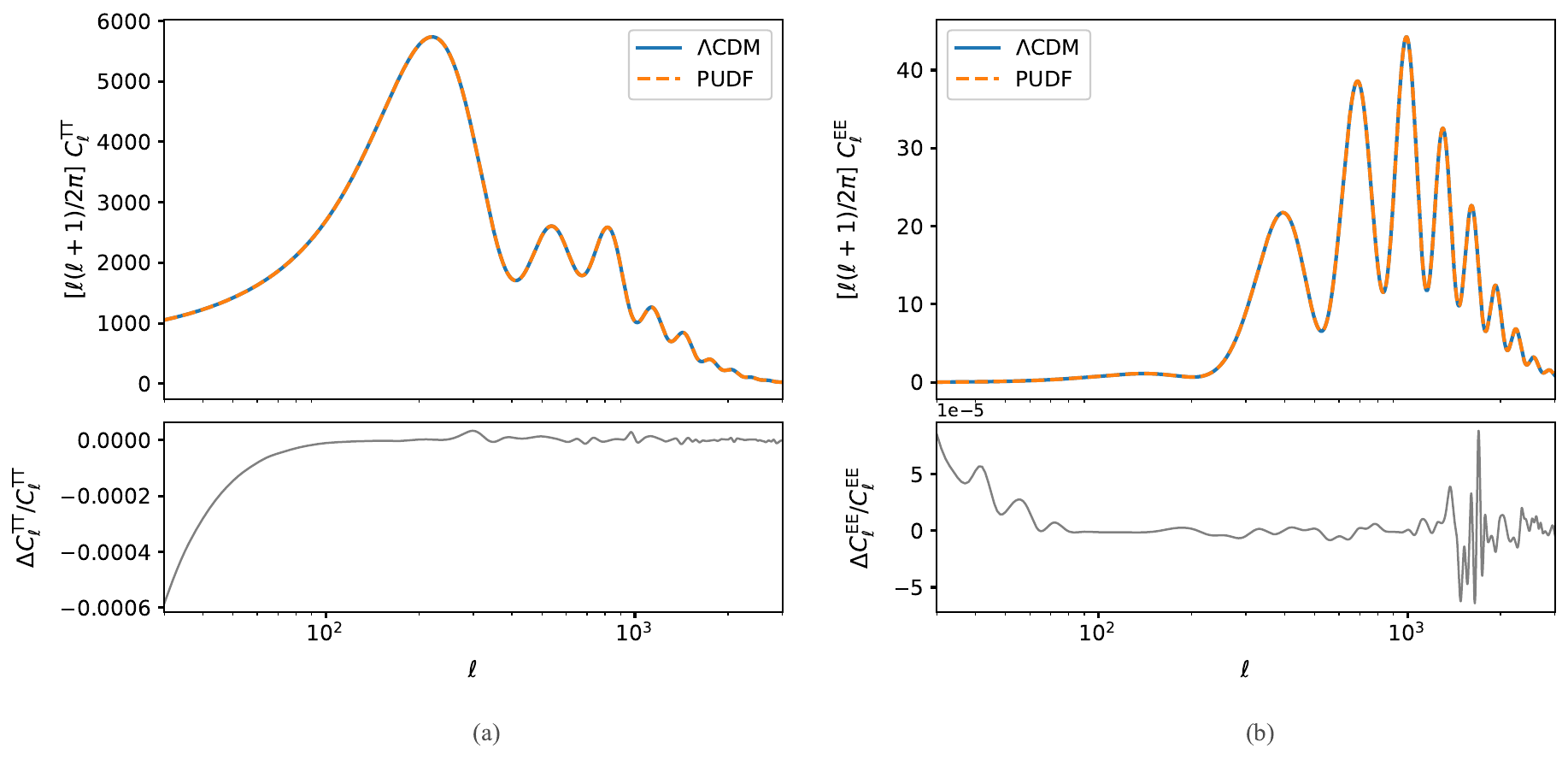}
\caption{Comparison of the primary CMB $TT$ and $EE$ power spectra of PUDF and $\Lambda$CDM when the matching condition~\eqref{eq:CMBmatchexact} is applied. The lower panels give the relative difference ($\Delta C_\ell^\mathrm{TT} = C^\mathrm{TT}_{\ell, \mathrm{PUDF}}-C^\mathrm{TT}_{\ell, \Lambda\mathrm{CDM}},\, 
\Delta C_\ell^\mathrm{EE}=C^\mathrm{EE}_{\ell, \mathrm{PUDF}}-C^\mathrm{EE}_{\ell, \Lambda\mathrm{CDM}}$).\label{fig:cl}}
\end{figure*}

We use Eq.~\eqref{eq:CMBmatchexact} to test the modified Boltzmann code CLASS in \citet{Wang:2024rus} and find an $O(10^{-3})$ relative difference between PUDF and $\Lambda$CDM primary CMB power spectra. Further investigation shows that this inconsistency is due to the usage of the subpackage HyRec, which contains a hard-coded $w_0w_a$CDM cosmology and therefore can be incompatible with modifications in CLASS. To fix this problem, we replace HyRec with the adapted version of RecFAST in CLASS, which reads cosmology from CLASS. The updated code agrees well with the theoretical expectation that once Eq.~\eqref{eq:CMBmatchexact} is satisfied, the relative difference in primary CMB power spectra of PUDF and $\Lambda$CDM does not exceed $O(10^{-4})$. Figure~\ref{fig:cl} shows an example where PUDF is matched to the Planck 2018 bestfit $\Lambda$CDM model~\citep{aghanim2020planck}. And the relevant corrections have been released in the associated erratum~\citep{Wang_2025}.

\subsection{Matching late-universe observables}

When $\beta$ becomes comparable to $p_{\rm age}$, which happens at $0\le z\lesssim 1$, the $\eta$ parameter becomes important and the unified dark fluid has negative pressure. We emphasize, however, that PUDF model should not be extended to the future $z<0$ regime, as $\eta$ is a phenomenological ``local'' parameter characterizing the average cosmic-acceleration in the recent epoch.

For a given $\Omega_m\rvert_{\rm \Lambda CDM}$, Eq.~\eqref{eq:CMBmatch} does not fix $p_{\rm age}$ and $\eta$. We may choose another constraint to match more observables between PUDF and $\Lambda$CDM. For instance, we may match the deceleration parameter $q_0 = -\frac{a\ddot a}{\dot a^2}$ in PUDF and $\Lambda$CDM. In the case of negligible $\Omega_\nu$ and $\Omega_\gamma$, the $q_0$ matching condition is
\begin{equation}
\left. \frac{4(1-\eta)}{9p_{\rm age}^2} \right\vert_{\rm PUDF} = \left.\Omega_m \right\vert_{\rm \Lambda CDM}. \label{eq:q0match}
\end{equation}
In the original work on PAge where only late universe observables were used, the primary-CMB matching condition~\eqref{eq:CMBmatch} was not considered. Instead, the age of the universe in the unit of $H_0$ was matched~\citep{Huang:2020mub}. Ignoring the radiation and neutrinos, the age matching condition is
\begin{equation}
\left. p_{\rm age} \right\vert_{\rm PUDF} = \left. \frac{2}{3\sqrt{1-\Omega_m}}\ln\frac{1+\sqrt{1-\Omega_m}}{\sqrt{\Omega_m}}\right\vert_{\rm \Lambda CDM}. \label{eq:agematch}
\end{equation}

\begin{figure*}
  \centering
\includegraphics[width=\textwidth]{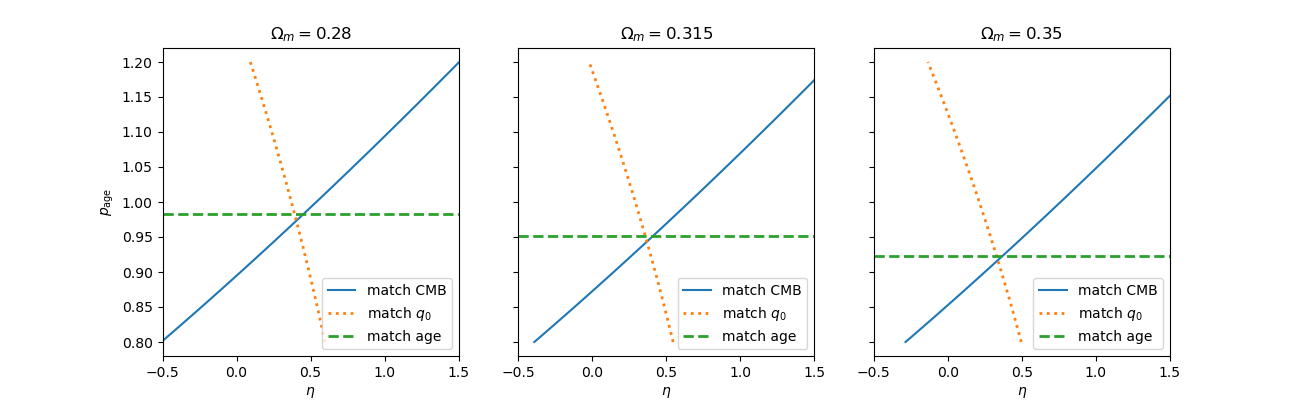}
\caption{The primary-CMB matching condition~\eqref{eq:CMBmatch}, $q_0$ matching condition~\eqref{eq:q0match} and age matching condition~\eqref{eq:agematch} for $\Omega_m=0.28$ (left panel), $\Omega_m=0.315$ (middle panel) and $\Omega_m=0.35$ (right panel), respectively. \label{fig:match}}
\end{figure*}

In Figure~\ref{fig:match} we plot the matching conditions for primary CMB, $q_0$ and age for a few representative $\Omega_m$ values. It is nontrivial to observe that the three conditions almost intersect at one point, where both early- and late-universe observables match well between PUDF and $\Lambda$CDM. It has been shown in \citet{Huang:2020mub} that BAO and SNe observables can be matched to percent-level accuracy between PAge and $\Lambda$CDM. 

\begin{figure}
  \centering
    \includegraphics[width=8cm]{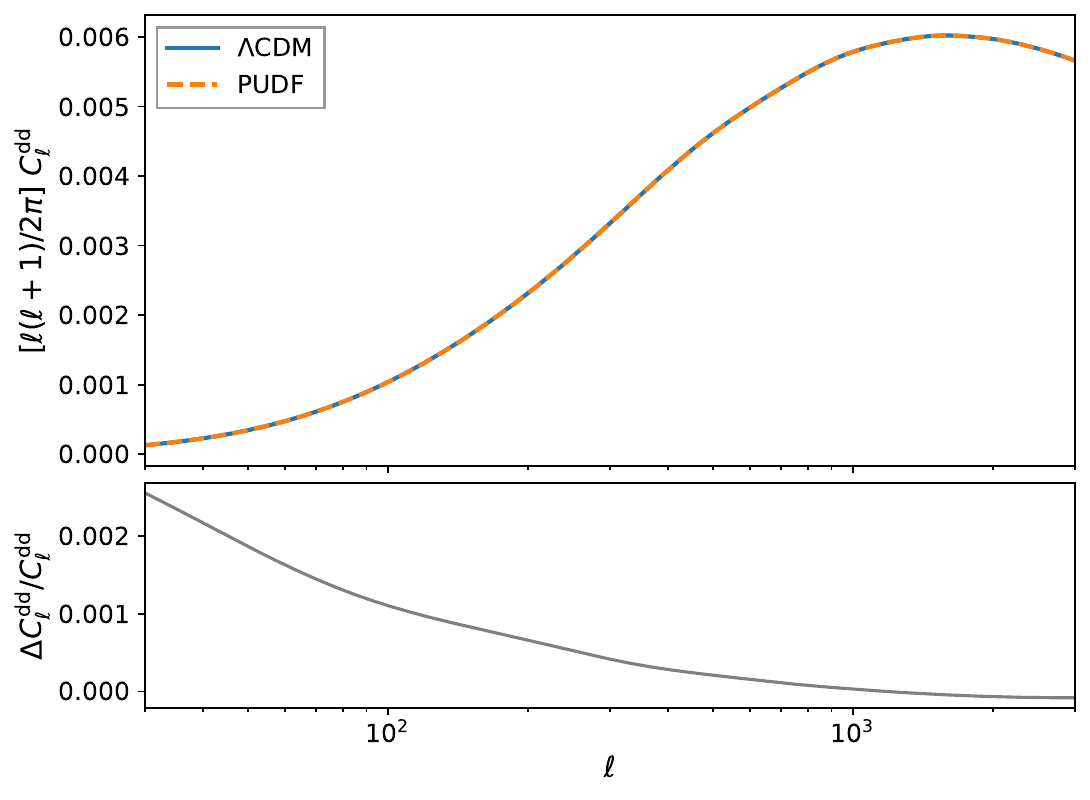}
    \caption{Comparison of CMB deflection power spectrum $C_{\ell}^{dd}$ of PUDF and $\Lambda$CDM when the primary-CMB matching condition~\eqref{eq:CMBmatchexact} and the $q_0$ matching condition~\eqref{eq:q0match} are applied. The lower panel gives the relative difference($\Delta C_\ell^\mathrm{dd} = C^\mathrm{dd}_{\ell, \mathrm{PUDF}}-C^\mathrm{dd}_{\ell, \Lambda\mathrm{CDM}}$).     \label{fig:cldd}}
\end{figure}

While the background evolution is matched between PUDF and $\Lambda$CDM, the abundance and equation of state of the unified dark fluid in PUDF are very different from those of dark matter in $\Lambda$CDM. We may expect very different density perturbations of the dark components in the two models. However, density perturbations of the dark components are not directly observable. What can be observed are the density perturbations of baryonic matter and the bending of the light due to gravitational lensing, both of which track the gravitational potential $\phi$ if anisotropic stress can be ignored. 
The linear growth of $\phi$ in general depends on the total density perturbation $\delta\rho_{\rm tot}$, the total pressure perturbation $\delta p_{\rm tot}$, and the expansion history of the universe~\citep{Weller_Lewis_2003}. On sub-horizon scales where the gauge-dependence of $\delta \rho_{\rm tot}$ and $\delta p_{\rm tot}$ can be ignored, we may use the Poisson equation to eliminate the dependence on $\delta\rho_{\rm tot}$~\citep{Weller_Lewis_2003}. Thus, in models such as PUDF and $\Lambda$CDM where the rest-frame pressure perturbations are assumed to be negligible, the evolution of $\phi$ on sub-horizon scales only depends on the expansion history of the universe. In other words, for background-matched PUDF and $\Lambda$CDM, the linear growth of gravitational potential is also approximately matched. This has been numerically verified in \citet{Wang:2024rus} where the baryon power spectrum in PUDF was shown to be similar to that in $\Lambda$CDM. In Figure~\ref{fig:cldd} we show that the CMB lensing deflection power spectrum in PUDF and $\Lambda$CDM are similar, too.  

\section{Parameter Inference}\label{sec:data}
In this section, we update the parameter inference for PUDF using the latest cosmological datasets.
Our analysis incorporates the CMB temperature, polarization, and lensing likelihoods~\citep{aghanim2020planck2, aghanim2020planck3}, the Dark Energy Survey 5-Year Type Ia supernova sample (DES5YR, \citet{DES:2024jxu}), BAO measurements from the DESI Data Release 2 (DESI DR2, \citet{DESI:2025zgx}), and CC determinations of the Hubble parameter~\citep{moresco20166}. The values of the inferred parameters are summarized in Table~\ref{tab:params}.
Compared to the corrected results reported in \citet{Wang_2025}, the updated PUDF parameters exhibit further deviations from their $\Lambda$CDM counterparts. 
Using the MCEvidence code~\citep{Heavens:2017afc, Liddle2007InformationCF}, we find an updated Bayesian evidence value of $\ln B_{\Lambda \mathrm{CDM}, \mathrm{PUDF}} = 5.50$, indicating strong support for $\Lambda$CDM.

\begin{table*}
\centering
\caption{Constraints on parameters from CMB+DESI DR2+DES5YR+CC (this work) and CMB+Pantheon Plus+DESI DR1+CC \citep{Wang_2025} are listed together for comparison.\label{tab:params}}
\begin{tabular}{cccc}
\hline\hline
\multirow{2}{*}{Parameters}  & \multirow{2}{*}{$\Lambda$CDM} & \multicolumn{2}{c}{PUDF} \\ 
\cmidrule(lr){3-4} 
           &                              & CMB+DESI DR2+DES5YR+CC & CMB+DESI DR1+Pantheon Plus+CC \\ 
\midrule
$100\Omega_b h^2$    & $2.248\pm 0.013$         & $2.258\pm 0.014$       & $2.251\pm 0.014$ \\
$\Omega_c h^2$       & $0.11856\pm 0.00074$     & -                      & - \\
$100\theta_*$        &  $1.04203\pm 0.00029$    & $1.04219\pm 0.00028$   & $1.04206\pm 0.00029$  \\
$\ln[10^{10}A_{s }]$ & $3.053\pm 0.015$         & $3.059\pm 0.015$       &  $3.053\pm 0.015$  \\
$n_s$                & $0.9688\pm 0.0035$       & $0.9727\pm 0.0038$     & $0.9697\pm 0.0039$   \\
$\tau_{\rm re}$      & $0.0595\pm 0.0075$       &$0.0637\pm 0.0078$     & $0.0599\pm 0.0077$  \\
$p_{\rm age}$        & -                        & $0.9620\pm 0.0064$     & $0.9619\pm 0.0073$ \\
$\eta$               &-                         & $0.421\pm 0.020$       & $0.428\pm 0.022$\\
$H_0$                & $68.05\pm 0.33$          &$68.15\pm 0.54$        & $68.14\pm 0.61$ \\
\midrule
$\ln B_{\rm \Lambda CDM, PUDF}$ &- &5.50  & 3.75\\
\hline
\end{tabular} 
\end{table*}

The differences in best-fit $\chi^2$ between PUDF and $\Lambda$CDM for various data combinations are listed in Table~\ref{tab:chisq}.
When CMB data is included, PUDF is less successful at tuning its parameters in a way that simultaneously fits both early- and late-universe observations compared to $\Lambda$CDM. However, as is shown in the last row of Table~\ref{tab:chisq}, when CMB data are replaced by a prior on $\Omega_bh^2$ from big bang nucleosynthesis (BBN) measurements~\citep{Pisanti:2007hk,adelberger2011solar,Aver2015,Cooke2018}, PUDF yields a slightly better fit, as its additional degree of freedom provides greater flexibility in adjusting the background expansion history than $\Lambda$CDM.
\begin{table}
\centering
\caption{Comparison of the relative minimum chi-square values, $\Delta\chi^2_{\rm min}$, of PUDF with respect to $\Lambda$CDM for various combinations of observational datasets \label{tab:chisq}}
\begin{tabular}{ll}
\hline
\hline
Datasets &  $\Delta\chi^2_{\rm min}$ \\
\midrule
CMB+DESI DR1+Pantheon Plus+CC &  $6.50$ \\
CMB+Pantheon Plus+CC &  $1.10$ \\
CMB+DESI DR1+CC & $0.58$  \\
BBN+DESI DR1+Pantheon Plus+CC & $-1.73$ \\
\hline
\end{tabular}
\end{table}

In summary, these results indicate that the difference between PUDF and $\Lambda$CDM becomes statistically significant only when CMB, BAO, and SNe datasets are jointly considered. 
However, it has been shown that the DESI BAO, CMB, and SNe measurements exhibit non-negligible mutual inconsistencies when interpreted within the $\Lambda$CDM framework~\citep{DESI:2024mwx,DESI:2025zgx, SPT25, Huang25bao}. This raises the concern that the strong Bayesian preference $\Delta\ln B = 5.50$ may be driven by unidentified systematics in the data rather than representing genuine evidence against PUDF. To assess this possibility, we perform the following two tests. 

\begin{figure*}
    \includegraphics[width=0.45\textwidth]{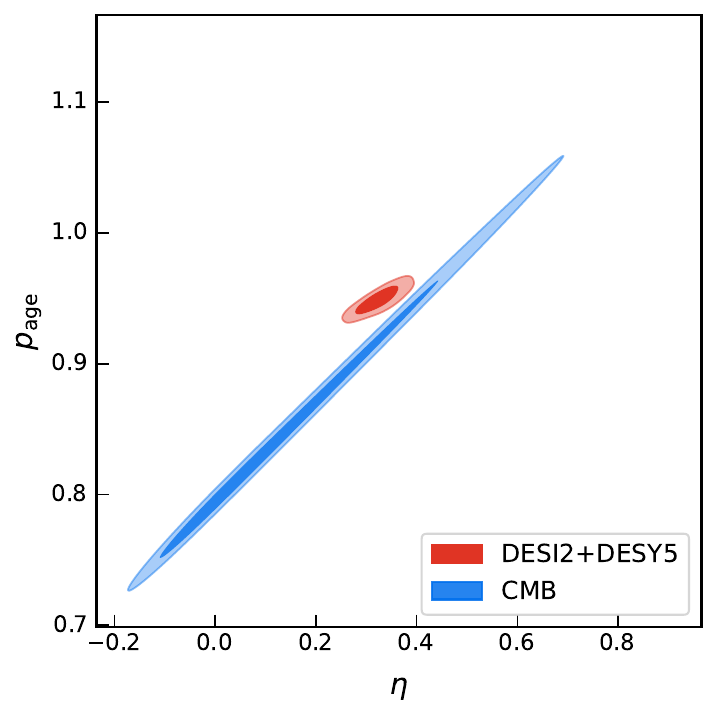}    
    \includegraphics[width=0.45\textwidth]{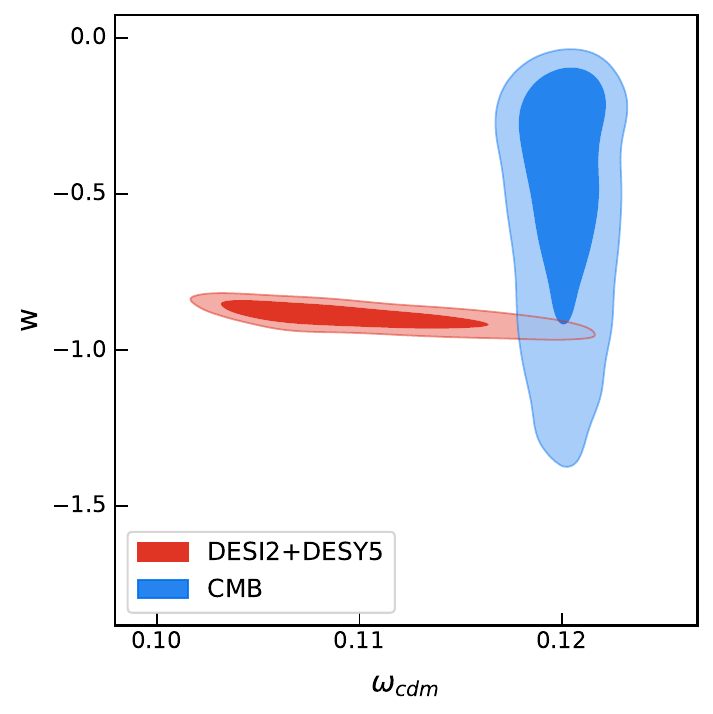}
    \caption{Two-dimensional marginalized contours at the $1\sigma$ and $2\sigma$ confidence levels for (a) the \{$\eta, p_{\rm age}$\} parameter space in PUDF and (b) the \{$\omega_{\rm cdm}\equiv \Omega_c h^2, w$\} parameter space in $w$CDM, derived from the CMB and DESI+DES5YR datasets.}    \label{fig:counter}
\end{figure*}

Firstly, we examine the consistency between datasets by plotting the two-dimensional marginalized posterior contours at the $1\sigma$ and $2\sigma$ confidence levels for the PUDF model using CMB alone and DESI DR2 + DES5YR data combination. The results are shown in Figure~\ref{fig:counter} (a). 
As illustrated, the two datasets favor markedly different regions in the \{$p_{\rm{age}}, \eta$\} parameter space, corresponding to a tension at the $4.38\sigma$ level.
For comparison, we also display in Figure~\ref{fig:counter} (b) the corresponding $1\sigma$ and $2\sigma$ contours for the $w$CDM model. In this case, the tension between the two datasets is reduced to $2.64\sigma$.
These results suggest that fully reconciling the discrepancy between early- and late-universe datasets with a single-parameter extension of $\Lambda$CDM, such as PUDF and $w$CDM, is a challenging task, although some specific models and parameterizations, such as $w$CDM, exhibit a relatively smaller tension. A statistically satisfactory fit to the combined dataset (DESI BAO + SNe + CMB) is likely to demand a two-parameter extension of the $\Lambda$CDM model, such as the Chevallier–Polarski–Linder (CPL) parametrization~\citep{Chevallier:2000qy,PhysRevLett.90.091301}.

Secondly, we generate mock data by replacing the central values of all observables in BAO, SNe, and CC with the theoretical predictions from the Planck 2018 best-fit $\Lambda$CDM model~\citep{aghanim2020planck}. When analyzed using these mock datasets, the best-fit $\chi^2$ difference between PUDF and $\Lambda$CDM reduces to $0.34$. This result indicates that the large $\chi^2$ difference observed in the real data would constitute a very rare fluctuation if $\Lambda$CDM were the underlying correct model.

Taken together, these findings suggest that the statistically significant preference for $\Lambda$CDM over PUDF observed in the real data ($\Delta\ln B = 5.50$) could be biased by the existing tension between the CMB and DESI BAO+SNe datasets. Alternatively, this may indicate that $\Lambda$CDM itself is not the correct model, pointing to the possible need for new physics beyond both PUDF and $\Lambda$CDM.

\section{Discussion and Conclusions} \label{sec:concludions}
In this study, we demonstrate that, within the PAge-like unified dark fluid model, both the background expansion history and the linear perturbations in the visible sector can be tuned to closely resemble those of $\Lambda$CDM. We give physical interpretation and derive the corresponding matching conditions for primary CMB and late-universe observations. These findings are in agreement with, e.g., \citet{Kou:2025yfr}.

We update the model comparison between PUDF and $\Lambda$CDM with the latest datasets including CMB, DESI DR2, DES5YR, and CC measurements. We find that the combined dataset favors $\Lambda$CDM over PUDF with a strong Bayesian evidence $\Delta\ln B = 5.5$. To determine whether this apparent preference stems from intrinsic limitations of the PUDF framework or from existing tensions among the CMB, DESI BAO, and SNe datasets, we assess dataset consistency. Under PUDF, a $4.38\sigma$ tension is found between CMB and DESI DR2+DES5YR, while it decreases to $2.64\sigma$ in $w$CDM. Simulations indicate that such discrepancies are very rare fluctuations if $\Lambda$CDM is indeed the true underlying model. In light of the unresolved tension in the current data, the joint statistical preference for $\Lambda$CDM over PUDF appears less robust. We anticipate that more precise future data will allow for a better distinction between PUDF and $\Lambda$CDM.

The key elements that make $\Lambda$CDM and PUDF match observations can be extended to other models with fluid-like dark components. Specifically, to match the present observational data, four conditions should be met: (i) the dark sector behaves as dust at high redshift ($z\gtrsim 1000$); (ii) the dark sector has negative pressure ($p/\rho <-1/3$) at low redshift; (iii) the shear perturbations of the dark sector are negligible; (iv) the pressure perturbations of the dark sector are negligible.  The most important lesson we have learned here is that the number of distinct dark components (1 in PUDF and 2 in $\Lambda$CDM) is not a key element for the model to match the visible part of the universe at the linear-perturbation level. To falsify PUDF (or $\Lambda$CDM), it is essential to go beyond linear perturbations. In the dark sector and on nonlinear scales, PUDF or in general a unified-dark-fluid model can be very different from $\Lambda$CDM. For instance, we are not sure if there can be unified-dark-fluid halos in the low-redshift universe, and if yes, whether their morphology is close to that in $\Lambda$CDM. The fluid description is a phenomenological large-scale approximation of an underlying fundamental theory which we have not yet specified. Given the tantalizing possibility of testing cosmology in the deep nonlinear regime with the future releases of DESI and other cosmological surveys, it would be an interesting direction to construct an underlying theory of PUDF and make predictions on nonlinear scales.

\printcredits



\begin{thebibliography}{77}
\expandafter\ifx\csname natexlab\endcsname\relax\def\natexlab#1{#1}\fi
\providecommand{\url}[1]{\texttt{#1}}
\providecommand{\href}[2]{#2}
\providecommand{\path}[1]{#1}
\providecommand{\DOIprefix}{doi:}
\providecommand{\ArXivprefix}{arXiv:}
\providecommand{\URLprefix}{URL: }
\providecommand{\Pubmedprefix}{pmid:}
\providecommand{\doi}[1]{\href{http://dx.doi.org/#1}{\path{#1}}}
\providecommand{\Pubmed}[1]{\href{pmid:#1}{\path{#1}}}
\providecommand{\bibinfo}[2]{#2}
\ifx\xfnm\relax \def\xfnm[#1]{\unskip,\space#1}\fi
\bibitem[{Abbott et~al.(2024)}]{DES:2024jxu}
\bibinfo{author}{Abbott, T.M.C.}, et~al. (\bibinfo{collaboration}{DES}),
  \bibinfo{year}{2024}.
\newblock \bibinfo{title}{{The Dark Energy Survey: Cosmology Results with
  {\ensuremath{\sim}}1500 New High-redshift Type Ia Supernovae Using the Full 5
  yr Data Set}}.
\newblock \bibinfo{journal}{Astrophys. J. Lett.} \bibinfo{volume}{973},
  \bibinfo{pages}{L14}.
\newblock \DOIprefix\doi{10.3847/2041-8213/ad6f9f},
  \href{http://arxiv.org/abs/2401.02929}{\tt arXiv:2401.02929}.
\bibitem[{{Abdul Karim} et~al.(2025){Abdul Karim}, {Aguilar}, {Ahlen}
  et~al.}]{DESI:2025zgx}
\bibinfo{author}{{Abdul Karim}, M.}, \bibinfo{author}{{Aguilar}, J.},
  \bibinfo{author}{{Ahlen}, S.}, et~al., \bibinfo{year}{2025}.
\newblock \bibinfo{title}{{DESI DR2 results. II. Measurements of baryon
  acoustic oscillations and cosmological constraints}}.
\newblock \bibinfo{journal}{\prd} \bibinfo{volume}{112},
  \bibinfo{pages}{083515}.
\newblock \DOIprefix\doi{10.1103/tr6y-kpc6},
  \href{http://arxiv.org/abs/2503.14738}{\tt arXiv:2503.14738}.
\bibitem[{Abdullah et~al.(2022)Abdullah, El-Zant and
  Ellithi}]{abdullah2022growth}
\bibinfo{author}{Abdullah, A.}, \bibinfo{author}{El-Zant, A.A.},
  \bibinfo{author}{Ellithi, A.}, \bibinfo{year}{2022}.
\newblock \bibinfo{title}{Growth of fluctuations in chaplygin gas cosmologies:
  A nonlinear jeans scale for unified dark matter}.
\newblock \bibinfo{journal}{Physical Review D} \bibinfo{volume}{106},
  \bibinfo{pages}{083524}.
\bibitem[{Adame et~al.(2025)}]{DESI:2024mwx}
\bibinfo{author}{Adame, A.G.}, et~al. (\bibinfo{collaboration}{DESI}),
  \bibinfo{year}{2025}.
\newblock \bibinfo{title}{{DESI 2024 VI: cosmological constraints from the
  measurements of baryon acoustic oscillations}}.
\newblock \bibinfo{journal}{JCAP} \bibinfo{volume}{02}, \bibinfo{pages}{021}.
\newblock \DOIprefix\doi{10.1088/1475-7516/2025/02/021},
  \href{http://arxiv.org/abs/2404.03002}{\tt arXiv:2404.03002}.
\bibitem[{Adelberger et~al.(2011)Adelberger, Garc{\'\i}a, Robertson, Snover,
  Balantekin, Heeger, Ramsey-Musolf, Bemmerer, Junghans, Bertulani
  et~al.}]{adelberger2011solar}
\bibinfo{author}{Adelberger, E.G.}, \bibinfo{author}{Garc{\'\i}a, A.},
  \bibinfo{author}{Robertson, R.H.}, \bibinfo{author}{Snover, K.},
  \bibinfo{author}{Balantekin, A.}, \bibinfo{author}{Heeger, K.},
  \bibinfo{author}{Ramsey-Musolf, M.}, \bibinfo{author}{Bemmerer, D.},
  \bibinfo{author}{Junghans, A.}, \bibinfo{author}{Bertulani, C.}, et~al.,
  \bibinfo{year}{2011}.
\newblock \bibinfo{title}{Solar fusion cross sections. ii. the pp chain and cno
  cycles}.
\newblock \bibinfo{journal}{Reviews of Modern Physics} \bibinfo{volume}{83},
  \bibinfo{pages}{195--245}.
\bibitem[{Aghanim et~al.(2020a)Aghanim, Akrami, Ashdown, Aumont, Baccigalupi,
  Ballardini, Banday, Barreiro, Bartolo, Basak et~al.}]{aghanim2020planck2}
\bibinfo{author}{Aghanim, N.}, \bibinfo{author}{Akrami, Y.},
  \bibinfo{author}{Ashdown, M.}, \bibinfo{author}{Aumont, J.},
  \bibinfo{author}{Baccigalupi, C.}, \bibinfo{author}{Ballardini, M.},
  \bibinfo{author}{Banday, A.J.}, \bibinfo{author}{Barreiro, R.},
  \bibinfo{author}{Bartolo, N.}, \bibinfo{author}{Basak, S.}, et~al.,
  \bibinfo{year}{2020}a.
\newblock \bibinfo{title}{Planck 2018 results-v. cmb power spectra and
  likelihoods}.
\newblock \bibinfo{journal}{Astronomy \& Astrophysics} \bibinfo{volume}{641},
  \bibinfo{pages}{A5}.
\bibitem[{Aghanim et~al.(2020b)Aghanim, Akrami, Ashdown, Aumont, Baccigalupi,
  Ballardini, Banday, Barreiro, Bartolo, Basak et~al.}]{aghanim2020planck}
\bibinfo{author}{Aghanim, N.}, \bibinfo{author}{Akrami, Y.},
  \bibinfo{author}{Ashdown, M.}, \bibinfo{author}{Aumont, J.},
  \bibinfo{author}{Baccigalupi, C.}, \bibinfo{author}{Ballardini, M.},
  \bibinfo{author}{Banday, A.J.}, \bibinfo{author}{Barreiro, R.},
  \bibinfo{author}{Bartolo, N.}, \bibinfo{author}{Basak, S.}, et~al.,
  \bibinfo{year}{2020}b.
\newblock \bibinfo{title}{Planck 2018 results-vi. cosmological parameters}.
\newblock \bibinfo{journal}{Astronomy \& Astrophysics} \bibinfo{volume}{641},
  \bibinfo{pages}{A6}.
\bibitem[{Aghanim et~al.(2020c)Aghanim, Akrami, Ashdown, Aumont, Baccigalupi,
  Ballardini, Banday, Barreiro, Bartolo, Basak et~al.}]{aghanim2020planck3}
\bibinfo{author}{Aghanim, N.}, \bibinfo{author}{Akrami, Y.},
  \bibinfo{author}{Ashdown, M.}, \bibinfo{author}{Aumont, J.},
  \bibinfo{author}{Baccigalupi, C.}, \bibinfo{author}{Ballardini, M.},
  \bibinfo{author}{Banday, A.J.}, \bibinfo{author}{Barreiro, R.},
  \bibinfo{author}{Bartolo, N.}, \bibinfo{author}{Basak, S.}, et~al.,
  \bibinfo{year}{2020}c.
\newblock \bibinfo{title}{Planck 2018 results-viii. gravitational lensing}.
\newblock \bibinfo{journal}{Astronomy \& Astrophysics} \bibinfo{volume}{641},
  \bibinfo{pages}{A8}.
\bibitem[{Armendariz-Picon et~al.(2001)Armendariz-Picon, Mukhanov and
  Steinhardt}]{armendariz2001essentials}
\bibinfo{author}{Armendariz-Picon, C.}, \bibinfo{author}{Mukhanov, V.},
  \bibinfo{author}{Steinhardt, P.J.}, \bibinfo{year}{2001}.
\newblock \bibinfo{title}{Essentials of k-essence}.
\newblock \bibinfo{journal}{Physical Review D} \bibinfo{volume}{63},
  \bibinfo{pages}{103510}.
\bibitem[{{Aver} et~al.(2015){Aver}, {Olive} and {Skillman}}]{Aver2015}
\bibinfo{author}{{Aver}, E.}, \bibinfo{author}{{Olive}, K.A.},
  \bibinfo{author}{{Skillman}, E.D.}, \bibinfo{year}{2015}.
\newblock \bibinfo{title}{{The effects of He I {\ensuremath{\lambda}}10830 on
  helium abundance determinations}}.
\newblock \bibinfo{journal}{\jcap} \bibinfo{volume}{2015},
  \bibinfo{pages}{011--011}.
\newblock \DOIprefix\doi{10.1088/1475-7516/2015/07/011},
  \href{http://arxiv.org/abs/1503.08146}{\tt arXiv:1503.08146}.
\bibitem[{Bento et~al.(2002)Bento, Bertolami and Sen}]{bento2002generalized}
\bibinfo{author}{Bento, M.}, \bibinfo{author}{Bertolami, O.},
  \bibinfo{author}{Sen, A.A.}, \bibinfo{year}{2002}.
\newblock \bibinfo{title}{Generalized chaplygin gas, accelerated expansion, and
  dark-energy-matter unification}.
\newblock \bibinfo{journal}{Physical Review D} \bibinfo{volume}{66},
  \bibinfo{pages}{043507}.
\bibitem[{Bertacca et~al.(2008)Bertacca, Bartolo, Diaferio and
  Matarrese}]{bertacca2008scalar}
\bibinfo{author}{Bertacca, D.}, \bibinfo{author}{Bartolo, N.},
  \bibinfo{author}{Diaferio, A.}, \bibinfo{author}{Matarrese, S.},
  \bibinfo{year}{2008}.
\newblock \bibinfo{title}{How the scalar field of unified dark matter models
  can cluster}.
\newblock \bibinfo{journal}{Journal of Cosmology and Astroparticle Physics}
  \bibinfo{volume}{2008}, \bibinfo{pages}{023}.
\bibitem[{Bertacca et~al.(2011)Bertacca, Bruni, Piattella and
  Pietrobon}]{bertacca2011unified}
\bibinfo{author}{Bertacca, D.}, \bibinfo{author}{Bruni, M.},
  \bibinfo{author}{Piattella, O.F.}, \bibinfo{author}{Pietrobon, D.},
  \bibinfo{year}{2011}.
\newblock \bibinfo{title}{Unified dark matter scalar field models with fast
  transition}.
\newblock \bibinfo{journal}{Journal of Cosmology and Astroparticle Physics}
  \bibinfo{volume}{2011}, \bibinfo{pages}{018}.
\bibitem[{Bili{\'c} et~al.(2002)Bili{\'c}, Tupper and
  Viollier}]{bilic2002unification}
\bibinfo{author}{Bili{\'c}, N.}, \bibinfo{author}{Tupper, G.B.},
  \bibinfo{author}{Viollier, R.D.}, \bibinfo{year}{2002}.
\newblock \bibinfo{title}{Unification of dark matter and dark energy: the
  inhomogeneous chaplygin gas}.
\newblock \bibinfo{journal}{Physics Letters B} \bibinfo{volume}{535},
  \bibinfo{pages}{17--21}.
\bibitem[{Blas et~al.(2011)Blas, Lesgourgues and Tram}]{Diego_Blas_2011}
\bibinfo{author}{Blas, D.}, \bibinfo{author}{Lesgourgues, J.},
  \bibinfo{author}{Tram, T.}, \bibinfo{year}{2011}.
\newblock \bibinfo{title}{The cosmic linear anisotropy solving system (class).
  part ii: Approximation schemes}.
\newblock \bibinfo{journal}{Journal of Cosmology and Astroparticle Physics}
  \bibinfo{volume}{2011}, \bibinfo{pages}{034–034}.
\newblock \URLprefix \url{http://dx.doi.org/10.1088/1475-7516/2011/07/034},
  \DOIprefix\doi{10.1088/1475-7516/2011/07/034}.
\bibitem[{{Brilenkov} et~al.(2013){Brilenkov}, {Eingorn}, {Jenkovszky} and
  {Zhuk}}]{Brilenkov2013}
\bibinfo{author}{{Brilenkov}, M.}, \bibinfo{author}{{Eingorn}, M.},
  \bibinfo{author}{{Jenkovszky}, L.}, \bibinfo{author}{{Zhuk}, A.},
  \bibinfo{year}{2013}.
\newblock \bibinfo{title}{{Dark matter and dark energy from quark bag model}}.
\newblock \bibinfo{journal}{\jcap} \bibinfo{volume}{2013},
  \bibinfo{pages}{002}.
\newblock \DOIprefix\doi{10.1088/1475-7516/2013/08/002},
  \href{http://arxiv.org/abs/1304.7521}{\tt arXiv:1304.7521}.
\bibitem[{{Camphuis} et~al.(2025){Camphuis}, {Quan}, {Balkenhol}
  et~al.}]{SPT25}
\bibinfo{author}{{Camphuis}, E.}, \bibinfo{author}{{Quan}, W.},
  \bibinfo{author}{{Balkenhol}, L.}, et~al., \bibinfo{year}{2025}.
\newblock \bibinfo{title}{{SPT-3G D1: CMB temperature and polarization power
  spectra and cosmology from 2019 and 2020 observations of the SPT-3G Main
  field}}.
\newblock \bibinfo{journal}{arXiv e-prints} ,
  \bibinfo{pages}{arXiv:2506.20707}\DOIprefix\doi{10.48550/arXiv.2506.20707},
  \href{http://arxiv.org/abs/2506.20707}{\tt arXiv:2506.20707}.
\bibitem[{Chavanis(2022)}]{chavanis2022k}
\bibinfo{author}{Chavanis, P.H.}, \bibinfo{year}{2022}.
\newblock \bibinfo{title}{K-essence lagrangians of polytropic and logotropic
  unified dark matter and dark energy models}.
\newblock \bibinfo{journal}{Astronomy} \bibinfo{volume}{1},
  \bibinfo{pages}{126--221}.
\bibitem[{Chevallier and Polarski(2001)}]{Chevallier:2000qy}
\bibinfo{author}{Chevallier, M.}, \bibinfo{author}{Polarski, D.},
  \bibinfo{year}{2001}.
\newblock \bibinfo{title}{{Accelerating universes with scaling dark matter}}.
\newblock \bibinfo{journal}{Int. J. Mod. Phys. D} \bibinfo{volume}{10},
  \bibinfo{pages}{213--224}.
\newblock \DOIprefix\doi{10.1142/S0218271801000822},
  \href{http://arxiv.org/abs/gr-qc/0009008}{\tt arXiv:gr-qc/0009008}.
\bibitem[{Colistete~Jr et~al.(2007)Colistete~Jr, Fabris, Tossa and
  Zimdahl}]{colistete2007bulk}
\bibinfo{author}{Colistete~Jr, R.}, \bibinfo{author}{Fabris, J.},
  \bibinfo{author}{Tossa, J.}, \bibinfo{author}{Zimdahl, W.},
  \bibinfo{year}{2007}.
\newblock \bibinfo{title}{Bulk viscous cosmology}.
\newblock \bibinfo{journal}{Physical Review D} \bibinfo{volume}{76},
  \bibinfo{pages}{103516}.
\bibitem[{{Cooke} and {Fumagalli}(2018)}]{Cooke2018}
\bibinfo{author}{{Cooke}, R.J.}, \bibinfo{author}{{Fumagalli}, M.},
  \bibinfo{year}{2018}.
\newblock \bibinfo{title}{{Measurement of the primordial helium abundance from
  the intergalactic medium}}.
\newblock \bibinfo{journal}{Nature Astronomy} \bibinfo{volume}{2},
  \bibinfo{pages}{957--961}.
\newblock \DOIprefix\doi{10.1038/s41550-018-0584-z},
  \href{http://arxiv.org/abs/1810.06561}{\tt arXiv:1810.06561}.
\bibitem[{{Cuzinatto} et~al.(2018){Cuzinatto}, {Medeiros}, {de Morais} and
  {Brandenberger}}]{Cuzinatto2018}
\bibinfo{author}{{Cuzinatto}, R.R.}, \bibinfo{author}{{Medeiros}, L.G.},
  \bibinfo{author}{{de Morais}, E.M.}, \bibinfo{author}{{Brandenberger}, R.H.},
  \bibinfo{year}{2018}.
\newblock \bibinfo{title}{{Analytic study of cosmological perturbations in a
  unified model of dark matter and dark energy with a sharp transition}}.
\newblock \bibinfo{journal}{Astroparticle Physics} \bibinfo{volume}{103},
  \bibinfo{pages}{98--107}.
\newblock \DOIprefix\doi{10.1016/j.astropartphys.2018.07.005},
  \href{http://arxiv.org/abs/1802.01232}{\tt arXiv:1802.01232}.
\bibitem[{{Das} and {Sur}(2023)}]{Das2023}
\bibinfo{author}{{Das}, S.}, \bibinfo{author}{{Sur}, S.}, \bibinfo{year}{2023}.
\newblock \bibinfo{title}{{A unified cosmological dark sector from a
  Bose-Einstein condensate}}.
\newblock \bibinfo{journal}{Physics of the Dark Universe} \bibinfo{volume}{42},
  \bibinfo{pages}{101331}.
\newblock \DOIprefix\doi{10.1016/j.dark.2023.101331},
  \href{http://arxiv.org/abs/2203.16402}{\tt arXiv:2203.16402}.
\bibitem[{{Davari} et~al.(2018){Davari}, {Malekjani} and
  {Artymowski}}]{Davari2018}
\bibinfo{author}{{Davari}, Z.}, \bibinfo{author}{{Malekjani}, M.},
  \bibinfo{author}{{Artymowski}, M.}, \bibinfo{year}{2018}.
\newblock \bibinfo{title}{{New parametrization for unified dark matter and dark
  energy}}.
\newblock \bibinfo{journal}{\prd} \bibinfo{volume}{97},
  \bibinfo{pages}{123525}.
\newblock \DOIprefix\doi{10.1103/PhysRevD.97.123525},
  \href{http://arxiv.org/abs/1805.11033}{\tt arXiv:1805.11033}.
\bibitem[{Dou et~al.(2011)Dou, Meng et~al.}]{dou2011bulk}
\bibinfo{author}{Dou, X.}, \bibinfo{author}{Meng, X.H.}, et~al.,
  \bibinfo{year}{2011}.
\newblock \bibinfo{title}{Bulk viscous cosmology: unified dark matter}.
\newblock \bibinfo{journal}{Advances in Astronomy} \bibinfo{volume}{2011},
  \bibinfo{pages}{829340}.
\bibitem[{Dunsby et~al.(2024)Dunsby, Luongo and Muccino}]{dunsby2024unifying}
\bibinfo{author}{Dunsby, P.K.}, \bibinfo{author}{Luongo, O.},
  \bibinfo{author}{Muccino, M.}, \bibinfo{year}{2024}.
\newblock \bibinfo{title}{Unifying the dark sector through a single matter
  fluid with nonzero pressure}.
\newblock \bibinfo{journal}{Physical Review D} \bibinfo{volume}{109},
  \bibinfo{pages}{023510}.
\bibitem[{Dutta et~al.(2018)Dutta, Khyllep, Saridakis, Tamanini and
  Vagnozzi}]{dutta2018cosmological}
\bibinfo{author}{Dutta, J.}, \bibinfo{author}{Khyllep, W.},
  \bibinfo{author}{Saridakis, E.N.}, \bibinfo{author}{Tamanini, N.},
  \bibinfo{author}{Vagnozzi, S.}, \bibinfo{year}{2018}.
\newblock \bibinfo{title}{Cosmological dynamics of mimetic gravity}.
\newblock \bibinfo{journal}{Journal of Cosmology and Astroparticle Physics}
  \bibinfo{volume}{2018}, \bibinfo{pages}{041}.
\bibitem[{Elkhateeb(2019)}]{elkhateeb2019dissipative}
\bibinfo{author}{Elkhateeb, E.}, \bibinfo{year}{2019}.
\newblock \bibinfo{title}{Dissipative unified dark fluid model}.
\newblock \bibinfo{journal}{International Journal of Modern Physics D}
  \bibinfo{volume}{28}, \bibinfo{pages}{1950110}.
\bibitem[{Elkhateeb and Hashim(2023)}]{elkhateeb2023dissipative}
\bibinfo{author}{Elkhateeb, E.A.}, \bibinfo{author}{Hashim, M.},
  \bibinfo{year}{2023}.
\newblock \bibinfo{title}{Dissipative unified dark fluid: Observational
  constraints}.
\newblock \bibinfo{journal}{Journal of High Energy Astrophysics}
  \bibinfo{volume}{37}, \bibinfo{pages}{3--14}.
\bibitem[{{Ferreira} and {Avelino}(2018)}]{Ferreira2018}
\bibinfo{author}{{Ferreira}, V.M.C.}, \bibinfo{author}{{Avelino}, P.P.},
  \bibinfo{year}{2018}.
\newblock \bibinfo{title}{{Extended family of generalized Chaplygin gas
  models}}.
\newblock \bibinfo{journal}{\prd} \bibinfo{volume}{98},
  \bibinfo{pages}{043515}.
\newblock \DOIprefix\doi{10.1103/PhysRevD.98.043515},
  \href{http://arxiv.org/abs/1807.04656}{\tt arXiv:1807.04656}.
\bibitem[{{Fortunato} et~al.(2025){Fortunato}, {Hip{\'o}lito-Ricaldi}, {Videla}
  and {Villanueva}}]{Fortunato2025}
\bibinfo{author}{{Fortunato}, J.A.S.}, \bibinfo{author}{{Hip{\'o}lito-Ricaldi},
  W.S.}, \bibinfo{author}{{Videla}, N.}, \bibinfo{author}{{Villanueva}, J.R.},
  \bibinfo{year}{2025}.
\newblock \bibinfo{title}{{Cosmic slowing down of acceleration with the
  Chaplygin{\textendash}Jacobi gas as a dark fluid?}}
\newblock \bibinfo{journal}{European Physical Journal C} \bibinfo{volume}{85},
  \bibinfo{pages}{274}.
\newblock \DOIprefix\doi{10.1140/epjc/s10052-025-13996-3},
  \href{http://arxiv.org/abs/2406.13132}{\tt arXiv:2406.13132}.
\bibitem[{{Frion} et~al.(2024){Frion}, {Camarena}, {Giani}, {Miranda},
  {Bertacca}, {Marra} and {Piattella}}]{frion2024bayesian}
\bibinfo{author}{{Frion}, E.}, \bibinfo{author}{{Camarena}, D.},
  \bibinfo{author}{{Giani}, L.}, \bibinfo{author}{{Miranda}, T.},
  \bibinfo{author}{{Bertacca}, D.}, \bibinfo{author}{{Marra}, V.},
  \bibinfo{author}{{Piattella}, O.F.}, \bibinfo{year}{2024}.
\newblock \bibinfo{title}{{Bayesian analysis of a Unified Dark Matter model
  with transition: can it alleviate the H0 tension?}}
\newblock \bibinfo{journal}{The Open Journal of Astrophysics}
  \bibinfo{volume}{7}, \bibinfo{pages}{17}.
\newblock \DOIprefix\doi{10.21105/astro.2307.06320},
  \href{http://arxiv.org/abs/2307.06320}{\tt arXiv:2307.06320}.
\bibitem[{{Gadbail} et~al.(2022){Gadbail}, {Arora} and {Sahoo}}]{Gadbail2022}
\bibinfo{author}{{Gadbail}, G.N.}, \bibinfo{author}{{Arora}, S.},
  \bibinfo{author}{{Sahoo}, P.K.}, \bibinfo{year}{2022}.
\newblock \bibinfo{title}{{Generalized Chaplygin gas and accelerating universe
  in f(Q , T) gravity}}.
\newblock \bibinfo{journal}{Physics of the Dark Universe} \bibinfo{volume}{37},
  \bibinfo{pages}{101074}.
\newblock \DOIprefix\doi{10.1016/j.dark.2022.101074},
  \href{http://arxiv.org/abs/2206.10336}{\tt arXiv:2206.10336}.
\bibitem[{{Gorini} et~al.(2008){Gorini}, {Kamenshchik}, {Moschella},
  {Piattella} and {Starobinsky}}]{Gorini2008}
\bibinfo{author}{{Gorini}, V.}, \bibinfo{author}{{Kamenshchik}, A.Y.},
  \bibinfo{author}{{Moschella}, U.}, \bibinfo{author}{{Piattella}, O.F.},
  \bibinfo{author}{{Starobinsky}, A.A.}, \bibinfo{year}{2008}.
\newblock \bibinfo{title}{{Gauge-invariant analysis of perturbations in
  Chaplygin gas unified models of dark matter and dark energy}}.
\newblock \bibinfo{journal}{\jcap} \bibinfo{volume}{2008},
  \bibinfo{pages}{016}.
\newblock \DOIprefix\doi{10.1088/1475-7516/2008/02/016},
  \href{http://arxiv.org/abs/0711.4242}{\tt arXiv:0711.4242}.
\bibitem[{Guendelman et~al.(2016)Guendelman, Nissimov and
  Pacheva}]{guendelman2016unified}
\bibinfo{author}{Guendelman, E.}, \bibinfo{author}{Nissimov, E.},
  \bibinfo{author}{Pacheva, S.}, \bibinfo{year}{2016}.
\newblock \bibinfo{title}{Unified dark energy and dust dark matter dual to
  quadratic purely kinetic k-essence}.
\newblock \bibinfo{journal}{The European Physical Journal C}
  \bibinfo{volume}{76}, \bibinfo{pages}{1--12}.
\bibitem[{{Hashim} and {El-Zant}(2025)}]{Hashim2025preprint}
\bibinfo{author}{{Hashim}, M.}, \bibinfo{author}{{El-Zant}, A.A.},
  \bibinfo{year}{2025}.
\newblock \bibinfo{title}{{Clustered unified dark sector cosmology: Background
  evolution and linear perturbations in light of observations}}.
\newblock \bibinfo{journal}{\prd} \bibinfo{volume}{111},
  \bibinfo{pages}{063522}.
\newblock \DOIprefix\doi{10.1103/PhysRevD.111.063522},
  \href{http://arxiv.org/abs/2502.01751}{\tt arXiv:2502.01751}.
\bibitem[{{Heavens} et~al.(2017){Heavens}, {Fantaye}, {Mootoovaloo}, {Eggers},
  {Hosenie}, {Kroon} and {Sellentin}}]{Heavens:2017afc}
\bibinfo{author}{{Heavens}, A.}, \bibinfo{author}{{Fantaye}, Y.},
  \bibinfo{author}{{Mootoovaloo}, A.}, \bibinfo{author}{{Eggers}, H.},
  \bibinfo{author}{{Hosenie}, Z.}, \bibinfo{author}{{Kroon}, S.},
  \bibinfo{author}{{Sellentin}, E.}, \bibinfo{year}{2017}.
\newblock \bibinfo{title}{{Marginal Likelihoods from Monte Carlo Markov
  Chains}}.
\newblock \bibinfo{journal}{arXiv e-prints} ,
  \bibinfo{pages}{arXiv:1704.03472}\DOIprefix\doi{10.48550/arXiv.1704.03472},
  \href{http://arxiv.org/abs/1704.03472}{\tt arXiv:1704.03472}.
\bibitem[{Henriques et~al.(2009)Henriques, Potting and
  S{\'a}}]{henriques2009unification}
\bibinfo{author}{Henriques, A.B.}, \bibinfo{author}{Potting, R.},
  \bibinfo{author}{S{\'a}, P.M.}, \bibinfo{year}{2009}.
\newblock \bibinfo{title}{Unification of inflation, dark energy, and dark
  matter within the salam-sezgin cosmological model}.
\newblock \bibinfo{journal}{Physical Review D} \bibinfo{volume}{79},
  \bibinfo{pages}{103522}.
\bibitem[{Huang et~al.(2021)Huang, Huang, Li and Zhou}]{Huang-2108-03959}
\bibinfo{author}{Huang, L.}, \bibinfo{author}{Huang, Z.}, \bibinfo{author}{Li,
  Z.}, \bibinfo{author}{Zhou, H.}, \bibinfo{year}{2021}.
\newblock \bibinfo{title}{{A More Accurate Parameterization based on cosmic Age
  (MAPAge)}}.
\newblock \bibinfo{journal}{{RAA}} \bibinfo{volume}{21}, \bibinfo{pages}{277}.
\newblock \DOIprefix\doi{10.1088/1674-4527/21/11/277},
  \href{http://arxiv.org/abs/2108.03959}{\tt arXiv:2108.03959}.
\bibitem[{Huang(2020)}]{Huang:2020mub}
\bibinfo{author}{Huang, Z.}, \bibinfo{year}{2020}.
\newblock \bibinfo{title}{{Supernova Magnitude Evolution and PAge
  Approximation}}.
\newblock \bibinfo{journal}{Astrophys. J. Lett.} \bibinfo{volume}{892},
  \bibinfo{pages}{L28}.
\newblock \DOIprefix\doi{10.3847/2041-8213/ab8011},
  \href{http://arxiv.org/abs/2001.06926}{\tt arXiv:2001.06926}.
\bibitem[{{Huang}(2025)}]{Huang25bao}
\bibinfo{author}{{Huang}, Z.}, \bibinfo{year}{2025}.
\newblock \bibinfo{title}{{Reionization optical depth and CMB-BAO tension in
  punctuated inflation}}.
\newblock \bibinfo{journal}{\mnras} \bibinfo{volume}{544},
  \bibinfo{pages}{2193--2199}.
\newblock \DOIprefix\doi{10.1093/mnras/staf1892},
  \href{http://arxiv.org/abs/2509.09086}{\tt arXiv:2509.09086}.
\bibitem[{{Joyce} et~al.(2015){Joyce}, {Jain}, {Khoury} and
  {Trodden}}]{Joyce2015}
\bibinfo{author}{{Joyce}, A.}, \bibinfo{author}{{Jain}, B.},
  \bibinfo{author}{{Khoury}, J.}, \bibinfo{author}{{Trodden}, M.},
  \bibinfo{year}{2015}.
\newblock \bibinfo{title}{{Beyond the cosmological standard model}}.
\newblock \bibinfo{journal}{\physrep} \bibinfo{volume}{568},
  \bibinfo{pages}{1--98}.
\newblock \DOIprefix\doi{10.1016/j.physrep.2014.12.002},
  \href{http://arxiv.org/abs/1407.0059}{\tt arXiv:1407.0059}.
\bibitem[{Kamenshchik et~al.(2001)Kamenshchik, Moschella and
  Pasquier}]{kamenshchik2001alternative}
\bibinfo{author}{Kamenshchik, A.}, \bibinfo{author}{Moschella, U.},
  \bibinfo{author}{Pasquier, V.}, \bibinfo{year}{2001}.
\newblock \bibinfo{title}{An alternative to quintessence}.
\newblock \bibinfo{journal}{Physics Letters B} \bibinfo{volume}{511},
  \bibinfo{pages}{265--268}.
\bibitem[{{Kleidis} and {Spyrou}(2015)}]{Kleidis2015}
\bibinfo{author}{{Kleidis}, K.}, \bibinfo{author}{{Spyrou}, N.K.},
  \bibinfo{year}{2015}.
\newblock \bibinfo{title}{{Polytropic dark matter flows illuminate dark energy
  and accelerated expansion}}.
\newblock \bibinfo{journal}{\aap} \bibinfo{volume}{576}, \bibinfo{pages}{A23}.
\newblock \DOIprefix\doi{10.1051/0004-6361/201424402},
  \href{http://arxiv.org/abs/1411.6789}{\tt arXiv:1411.6789}.
\bibitem[{{Kou} and {Lewis}(2026)}]{Kou:2025yfr}
\bibinfo{author}{{Kou}, R.}, \bibinfo{author}{{Lewis}, A.},
  \bibinfo{year}{2026}.
\newblock \bibinfo{title}{{Unified dark fluid with null sound speed as an
  alternative to phantom dark energy}}.
\newblock \bibinfo{journal}{\jcap} \bibinfo{volume}{2026},
  \bibinfo{pages}{014}.
\newblock \DOIprefix\doi{10.1088/1475-7516/2026/01/014},
  \href{http://arxiv.org/abs/2509.16155}{\tt arXiv:2509.16155}.
\bibitem[{Koutsoumbas et~al.(2018)Koutsoumbas, Ntrekis, Papantonopoulos and
  Saridakis}]{koutsoumbas2018unification}
\bibinfo{author}{Koutsoumbas, G.}, \bibinfo{author}{Ntrekis, K.},
  \bibinfo{author}{Papantonopoulos, E.}, \bibinfo{author}{Saridakis, E.N.},
  \bibinfo{year}{2018}.
\newblock \bibinfo{title}{Unification of dark matter-dark energy in generalized
  galileon theories}.
\newblock \bibinfo{journal}{Journal of Cosmology and Astroparticle Physics}
  \bibinfo{volume}{2018}, \bibinfo{pages}{003}.
\bibitem[{{Kumar} and {Sen}(2014)}]{Kumar2014}
\bibinfo{author}{{Kumar}, S.}, \bibinfo{author}{{Sen}, A.A.},
  \bibinfo{year}{2014}.
\newblock \bibinfo{title}{{Clustering GCG: a viable option for unified dark
  matter-dark energy?}}
\newblock \bibinfo{journal}{\jcap} \bibinfo{volume}{2014},
  \bibinfo{pages}{036--036}.
\newblock \DOIprefix\doi{10.1088/1475-7516/2014/10/036},
  \href{http://arxiv.org/abs/1405.5688}{\tt arXiv:1405.5688}.
\bibitem[{{Li} and {Xu}(2013)}]{li2013viscous}
\bibinfo{author}{{Li}, W.}, \bibinfo{author}{{Xu}, L.}, \bibinfo{year}{2013}.
\newblock \bibinfo{title}{{Viscous generalized Chaplygin gas as a unified dark
  fluid}}.
\newblock \bibinfo{journal}{European Physical Journal C} \bibinfo{volume}{73},
  \bibinfo{pages}{2471}.
\newblock \DOIprefix\doi{10.1140/epjc/s10052-013-2471-1}.
\bibitem[{Liddle(2007)}]{Liddle2007InformationCF}
\bibinfo{author}{Liddle, A.R.}, \bibinfo{year}{2007}.
\newblock \bibinfo{title}{Information criteria for astrophysical model
  selection}.
\newblock \bibinfo{journal}{Monthly Notices of the Royal Astronomical Society:
  Letters} \bibinfo{volume}{377}.
\newblock \URLprefix \url{https://api.semanticscholar.org/CorpusID:2884450}.
\bibitem[{Liddle and Ure{\~n}a-L{\'o}pez(2006)}]{liddle2006inflation}
\bibinfo{author}{Liddle, A.R.}, \bibinfo{author}{Ure{\~n}a-L{\'o}pez, L.A.},
  \bibinfo{year}{2006}.
\newblock \bibinfo{title}{Inflation, dark matter, and dark energy in the string
  landscape}.
\newblock \bibinfo{journal}{Physical review letters} \bibinfo{volume}{97},
  \bibinfo{pages}{161301}.
\bibitem[{Linder(2003)}]{PhysRevLett.90.091301}
\bibinfo{author}{Linder, E.V.}, \bibinfo{year}{2003}.
\newblock \bibinfo{title}{Exploring the expansion history of the universe}.
\newblock \bibinfo{journal}{Phys. Rev. Lett.} \bibinfo{volume}{90},
  \bibinfo{pages}{091301}.
\newblock \URLprefix
  \url{https://link.aps.org/doi/10.1103/PhysRevLett.90.091301},
  \DOIprefix\doi{10.1103/PhysRevLett.90.091301}.
\bibitem[{{Lu} et~al.(2015){Lu}, {Geng}, {Xu}, {Wu} and {Liu}}]{lu2015}
\bibinfo{author}{{Lu}, J.}, \bibinfo{author}{{Geng}, D.},
  \bibinfo{author}{{Xu}, L.}, \bibinfo{author}{{Wu}, Y.},
  \bibinfo{author}{{Liu}, M.}, \bibinfo{year}{2015}.
\newblock \bibinfo{title}{{Reduced modified Chaplygin gas cosmology}}.
\newblock \bibinfo{journal}{Journal of High Energy Physics}
  \bibinfo{volume}{2015}, \bibinfo{pages}{71}.
\newblock \DOIprefix\doi{10.1007/JHEP02(2015)071},
  \href{http://arxiv.org/abs/1312.0779}{\tt arXiv:1312.0779}.
\bibitem[{Luo et~al.(2020)Luo, Huang, Qian and Huang}]{luo2020reaffirming}
\bibinfo{author}{Luo, X.}, \bibinfo{author}{Huang, Z.}, \bibinfo{author}{Qian,
  Q.}, \bibinfo{author}{Huang, L.}, \bibinfo{year}{2020}.
\newblock \bibinfo{title}{Reaffirming the cosmic acceleration without
  supernovae and the cosmic microwave background}.
\newblock \bibinfo{journal}{The Astrophysical Journal} \bibinfo{volume}{905},
  \bibinfo{pages}{53}.
\bibitem[{Mandal and Biswas(2024)}]{mandal2024dynamical}
\bibinfo{author}{Mandal, G.}, \bibinfo{author}{Biswas, S.K.},
  \bibinfo{year}{2024}.
\newblock \bibinfo{title}{Dynamical systems analysis of a cosmological model
  with interacting umami chaplygin fluid in adiabatic particle creation
  mechanism: Some bouncing features}.
\newblock \bibinfo{journal}{arXiv preprint arXiv:2403.13028} .
\bibitem[{{Martin}(2012)}]{Jerome2012}
\bibinfo{author}{{Martin}, J.}, \bibinfo{year}{2012}.
\newblock \bibinfo{title}{{Everything you always wanted to know about the
  cosmological constant problem (but were afraid to ask)}}.
\newblock \bibinfo{journal}{Comptes Rendus Physique} \bibinfo{volume}{13},
  \bibinfo{pages}{566--665}.
\newblock \DOIprefix\doi{10.1016/j.crhy.2012.04.008},
  \href{http://arxiv.org/abs/1205.3365}{\tt arXiv:1205.3365}.
\bibitem[{Mishra and Sahni(2021)}]{mishra2021unifying}
\bibinfo{author}{Mishra, S.S.}, \bibinfo{author}{Sahni, V.},
  \bibinfo{year}{2021}.
\newblock \bibinfo{title}{Unifying dark matter and dark energy with
  non-canonical scalars}.
\newblock \bibinfo{journal}{The European Physical Journal C}
  \bibinfo{volume}{81}, \bibinfo{pages}{1--10}.
\bibitem[{Moresco et~al.(2016)Moresco, Pozzetti, Cimatti, Jimenez, Maraston,
  Verde, Thomas, Citro, Tojeiro and Wilkinson}]{moresco20166}
\bibinfo{author}{Moresco, M.}, \bibinfo{author}{Pozzetti, L.},
  \bibinfo{author}{Cimatti, A.}, \bibinfo{author}{Jimenez, R.},
  \bibinfo{author}{Maraston, C.}, \bibinfo{author}{Verde, L.},
  \bibinfo{author}{Thomas, D.}, \bibinfo{author}{Citro, A.},
  \bibinfo{author}{Tojeiro, R.}, \bibinfo{author}{Wilkinson, D.},
  \bibinfo{year}{2016}.
\newblock \bibinfo{title}{A 6\% measurement of the hubble parameter at z~ 0.45:
  direct evidence of the epoch of cosmic re-acceleration}.
\newblock \bibinfo{journal}{Journal of Cosmology and Astroparticle Physics}
  \bibinfo{volume}{2016}, \bibinfo{pages}{014}.
\bibitem[{{Perlmutter} et~al.(1999){Perlmutter}, {Aldering}, {Goldhaber},
  {Knop}, {Nugent}, {Castro}, {Deustua}, {Fabbro}, {Goobar}, {Groom}, {Hook},
  {Kim}, {Kim}, {Lee}, {Nunes}, {Pain}, {Pennypacker}, {Quimby}, {Lidman},
  {Ellis}, {Irwin}, {McMahon}, {Ruiz-Lapuente}, {Walton}, {Schaefer}, {Boyle},
  {Filippenko}, {Matheson}, {Fruchter}, {Panagia}, {Newberg}, {Couch} and
  {Project}}]{1999ApJ...517..565P}
\bibinfo{author}{{Perlmutter}, S.}, \bibinfo{author}{{Aldering}, G.},
  \bibinfo{author}{{Goldhaber}, G.}, \bibinfo{author}{{Knop}, R.A.},
  \bibinfo{author}{{Nugent}, P.}, \bibinfo{author}{{Castro}, P.G.},
  \bibinfo{author}{{Deustua}, S.}, \bibinfo{author}{{Fabbro}, S.},
  \bibinfo{author}{{Goobar}, A.}, \bibinfo{author}{{Groom}, D.E.},
  \bibinfo{author}{{Hook}, I.M.}, \bibinfo{author}{{Kim}, A.G.},
  \bibinfo{author}{{Kim}, M.Y.}, \bibinfo{author}{{Lee}, J.C.},
  \bibinfo{author}{{Nunes}, N.J.}, \bibinfo{author}{{Pain}, R.},
  \bibinfo{author}{{Pennypacker}, C.R.}, \bibinfo{author}{{Quimby}, R.},
  \bibinfo{author}{{Lidman}, C.}, \bibinfo{author}{{Ellis}, R.S.},
  \bibinfo{author}{{Irwin}, M.}, \bibinfo{author}{{McMahon}, R.G.},
  \bibinfo{author}{{Ruiz-Lapuente}, P.}, \bibinfo{author}{{Walton}, N.},
  \bibinfo{author}{{Schaefer}, B.}, \bibinfo{author}{{Boyle}, B.J.},
  \bibinfo{author}{{Filippenko}, A.V.}, \bibinfo{author}{{Matheson}, T.},
  \bibinfo{author}{{Fruchter}, A.S.}, \bibinfo{author}{{Panagia}, N.},
  \bibinfo{author}{{Newberg}, H.J.M.}, \bibinfo{author}{{Couch}, W.J.},
  \bibinfo{author}{{Project}, T.S.C.}, \bibinfo{year}{1999}.
\newblock \bibinfo{title}{{Measurements of {\ensuremath{\Omega}} and
  {\ensuremath{\Lambda}} from 42 High-Redshift Supernovae}}.
\newblock \bibinfo{journal}{\apj} \bibinfo{volume}{517},
  \bibinfo{pages}{565--586}.
\newblock \DOIprefix\doi{10.1086/307221},
  \href{http://arxiv.org/abs/astro-ph/9812133}{\tt arXiv:astro-ph/9812133}.
\bibitem[{Pisanti et~al.(2008)Pisanti, Cirillo, Esposito, Iocco, Mangano, Miele
  and Serpico}]{Pisanti:2007hk}
\bibinfo{author}{Pisanti, O.}, \bibinfo{author}{Cirillo, A.},
  \bibinfo{author}{Esposito, S.}, \bibinfo{author}{Iocco, F.},
  \bibinfo{author}{Mangano, G.}, \bibinfo{author}{Miele, G.},
  \bibinfo{author}{Serpico, P.D.}, \bibinfo{year}{2008}.
\newblock \bibinfo{title}{{PArthENoPE: Public Algorithm Evaluating the
  Nucleosynthesis of Primordial Elements}}.
\newblock \bibinfo{journal}{Comput. Phys. Commun.} \bibinfo{volume}{178},
  \bibinfo{pages}{956--971}.
\newblock \DOIprefix\doi{10.1016/j.cpc.2008.02.015},
  \href{http://arxiv.org/abs/0705.0290}{\tt arXiv:0705.0290}.
\bibitem[{{Quiros} et~al.(2025){Quiros}, {Gonzalez}, {Nucamendi}, {De Arcia}
  and {Horta Rangel}}]{Quiros2025preprint}
\bibinfo{author}{{Quiros}, I.}, \bibinfo{author}{{Gonzalez}, T.},
  \bibinfo{author}{{Nucamendi}, U.}, \bibinfo{author}{{De Arcia}, R.},
  \bibinfo{author}{{Horta Rangel}, F.A.}, \bibinfo{year}{2025}.
\newblock \bibinfo{title}{{Revisiting purely kinetic k-essence}}.
\newblock \bibinfo{journal}{arXiv e-prints} ,
  \bibinfo{pages}{arXiv:2501.14177}\DOIprefix\doi{10.48550/arXiv.2501.14177},
  \href{http://arxiv.org/abs/2501.14177}{\tt arXiv:2501.14177}.
\bibitem[{{Radicella} and {Pav{\'o}n}(2014)}]{Radicella2014}
\bibinfo{author}{{Radicella}, N.}, \bibinfo{author}{{Pav{\'o}n}, D.},
  \bibinfo{year}{2014}.
\newblock \bibinfo{title}{{Thermodynamics of the unified dark fluid with fast
  transition}}.
\newblock \bibinfo{journal}{\prd} \bibinfo{volume}{89},
  \bibinfo{pages}{067302}.
\newblock \DOIprefix\doi{10.1103/PhysRevD.89.067302},
  \href{http://arxiv.org/abs/1403.2601}{\tt arXiv:1403.2601}.
\bibitem[{{Riess} et~al.(1998){Riess}, {Filippenko}, {Challis}, {Clocchiatti},
  {Diercks}, {Garnavich}, {Gilliland}, {Hogan}, {Jha}, {Kirshner},
  {Leibundgut}, {Phillips}, {Reiss}, {Schmidt}, {Schommer}, {Smith},
  {Spyromilio}, {Stubbs}, {Suntzeff} and {Tonry}}]{1998AJ....116.1009R}
\bibinfo{author}{{Riess}, A.G.}, \bibinfo{author}{{Filippenko}, A.V.},
  \bibinfo{author}{{Challis}, P.}, \bibinfo{author}{{Clocchiatti}, A.},
  \bibinfo{author}{{Diercks}, A.}, \bibinfo{author}{{Garnavich}, P.M.},
  \bibinfo{author}{{Gilliland}, R.L.}, \bibinfo{author}{{Hogan}, C.J.},
  \bibinfo{author}{{Jha}, S.}, \bibinfo{author}{{Kirshner}, R.P.},
  \bibinfo{author}{{Leibundgut}, B.}, \bibinfo{author}{{Phillips}, M.M.},
  \bibinfo{author}{{Reiss}, D.}, \bibinfo{author}{{Schmidt}, B.P.},
  \bibinfo{author}{{Schommer}, R.A.}, \bibinfo{author}{{Smith}, R.C.},
  \bibinfo{author}{{Spyromilio}, J.}, \bibinfo{author}{{Stubbs}, C.},
  \bibinfo{author}{{Suntzeff}, N.B.}, \bibinfo{author}{{Tonry}, J.},
  \bibinfo{year}{1998}.
\newblock \bibinfo{title}{{Observational Evidence from Supernovae for an
  Accelerating Universe and a Cosmological Constant}}.
\newblock \bibinfo{journal}{\aj} \bibinfo{volume}{116},
  \bibinfo{pages}{1009--1038}.
\newblock \DOIprefix\doi{10.1086/300499},
  \href{http://arxiv.org/abs/astro-ph/9805201}{\tt arXiv:astro-ph/9805201}.
\bibitem[{S{\'a}(2020)}]{sa2020unified}
\bibinfo{author}{S{\'a}, P.M.}, \bibinfo{year}{2020}.
\newblock \bibinfo{title}{Unified description of dark energy and dark matter
  within the generalized hybrid metric-palatini theory of gravity}.
\newblock \bibinfo{journal}{Universe} \bibinfo{volume}{6}, \bibinfo{pages}{78}.
\bibitem[{{Sahni} and {Sen}(2017)}]{Sahni2017}
\bibinfo{author}{{Sahni}, V.}, \bibinfo{author}{{Sen}, A.A.},
  \bibinfo{year}{2017}.
\newblock \bibinfo{title}{{A new recipe for $\Lambda$CDM}}.
\newblock \bibinfo{journal}{European Physical Journal C} \bibinfo{volume}{77},
  \bibinfo{pages}{225}.
\newblock \DOIprefix\doi{10.1140/epjc/s10052-017-4796-7},
  \href{http://arxiv.org/abs/1510.09010}{\tt arXiv:1510.09010}.
\bibitem[{{Sandvik} et~al.(2004){Sandvik}, {Tegmark}, {Zaldarriaga} and
  {Waga}}]{Sandvik2004}
\bibinfo{author}{{Sandvik}, H.B.}, \bibinfo{author}{{Tegmark}, M.},
  \bibinfo{author}{{Zaldarriaga}, M.}, \bibinfo{author}{{Waga}, I.},
  \bibinfo{year}{2004}.
\newblock \bibinfo{title}{{The end of unified dark matter?}}
\newblock \bibinfo{journal}{\prd} \bibinfo{volume}{69},
  \bibinfo{pages}{123524}.
\newblock \DOIprefix\doi{10.1103/PhysRevD.69.123524},
  \href{http://arxiv.org/abs/astro-ph/0212114}{\tt arXiv:astro-ph/0212114}.
\bibitem[{Scherrer(2004)}]{scherrer2004purely}
\bibinfo{author}{Scherrer, R.J.}, \bibinfo{year}{2004}.
\newblock \bibinfo{title}{Purely kinetic k essence as unified dark matter}.
\newblock \bibinfo{journal}{Physical review letters} \bibinfo{volume}{93},
  \bibinfo{pages}{011301}.
\bibitem[{{Shukla} et~al.(2025){Shukla}, {Tiwari}, {Beesham} and
  {Sofuo{\u{g}}lu}}]{Shukla2025}
\bibinfo{author}{{Shukla}, B.K.}, \bibinfo{author}{{Tiwari}, R.K.},
  \bibinfo{author}{{Beesham}, A.}, \bibinfo{author}{{Sofuo{\u{g}}lu}, D.},
  \bibinfo{year}{2025}.
\newblock \bibinfo{title}{Modified chaplygin gas solutions of f(q) theory of
  gravity}.
\newblock \bibinfo{journal}{\na} \bibinfo{volume}{117},
  \bibinfo{pages}{102355}.
\newblock \DOIprefix\doi{10.1016/j.newast.2025.102355}.
\bibitem[{{Tripathy} et~al.(2015){Tripathy}, {Behera} and
  {Mishra}}]{Tripathy2015}
\bibinfo{author}{{Tripathy}, S.K.}, \bibinfo{author}{{Behera}, D.},
  \bibinfo{author}{{Mishra}, B.}, \bibinfo{year}{2015}.
\newblock \bibinfo{title}{{Unified dark fluid in Brans{\textendash}Dicke
  theory}}.
\newblock \bibinfo{journal}{European Physical Journal C} \bibinfo{volume}{75},
  \bibinfo{pages}{149}.
\newblock \DOIprefix\doi{10.1140/epjc/s10052-015-3371-3},
  \href{http://arxiv.org/abs/1410.3156}{\tt arXiv:1410.3156}.
\bibitem[{{Tripathy} et~al.(2020){Tripathy}, {Pradhan}, {Naik}, {Behera} and
  {Mishra}}]{Tripathy2020}
\bibinfo{author}{{Tripathy}, S.K.}, \bibinfo{author}{{Pradhan}, S.K.},
  \bibinfo{author}{{Naik}, Z.}, \bibinfo{author}{{Behera}, D.},
  \bibinfo{author}{{Mishra}, B.}, \bibinfo{year}{2020}.
\newblock \bibinfo{title}{{Unified dark fluid and cosmic transit models in
  Brans-Dicke theory}}.
\newblock \bibinfo{journal}{Physics of the Dark Universe} \bibinfo{volume}{30},
  \bibinfo{pages}{100722}.
\newblock \DOIprefix\doi{10.1016/j.dark.2020.100722},
  \href{http://arxiv.org/abs/2004.01027}{\tt arXiv:2004.01027}.
\bibitem[{Wang et~al.(2024)Wang, Huang, Yao, Liu, Huang and Su}]{Wang:2024rus}
\bibinfo{author}{Wang, J.}, \bibinfo{author}{Huang, Z.}, \bibinfo{author}{Yao,
  Y.}, \bibinfo{author}{Liu, J.}, \bibinfo{author}{Huang, L.},
  \bibinfo{author}{Su, Y.}, \bibinfo{year}{2024}.
\newblock \bibinfo{title}{{A PAge-like Unified Dark Fluid model}}.
\newblock \bibinfo{journal}{JCAP} \bibinfo{volume}{09}, \bibinfo{pages}{053}.
\newblock \DOIprefix\doi{10.1088/1475-7516/2024/09/053},
  \href{http://arxiv.org/abs/2405.05798}{\tt arXiv:2405.05798}.
\bibitem[{Wang et~al.(2025)Wang, Huang, Yao, Liu, Huang and Su}]{Wang_2025}
\bibinfo{author}{Wang, J.}, \bibinfo{author}{Huang, Z.}, \bibinfo{author}{Yao,
  Y.}, \bibinfo{author}{Liu, J.}, \bibinfo{author}{Huang, L.},
  \bibinfo{author}{Su, Y.}, \bibinfo{year}{2025}.
\newblock \bibinfo{title}{Erratum: A page-like unified dark fluid model}.
\newblock \bibinfo{journal}{Journal of Cosmology and Astroparticle Physics}
  \bibinfo{volume}{2025}, \bibinfo{pages}{E01}.
\newblock \URLprefix \url{https://doi.org/10.1088/1475-7516/2025/09/E01},
  \DOIprefix\doi{10.1088/1475-7516/2025/09/E01}.
\bibitem[{{Weinberg}(1989)}]{Weinberg1989}
\bibinfo{author}{{Weinberg}, S.}, \bibinfo{year}{1989}.
\newblock \bibinfo{title}{{The cosmological constant problem}}.
\newblock \bibinfo{journal}{Reviews of Modern Physics} \bibinfo{volume}{61},
  \bibinfo{pages}{1--23}.
\newblock \DOIprefix\doi{10.1103/RevModPhys.61.1}.
\bibitem[{{Weller} and {Lewis}(2003)}]{Weller_Lewis_2003}
\bibinfo{author}{{Weller}, J.}, \bibinfo{author}{{Lewis}, A.M.},
  \bibinfo{year}{2003}.
\newblock \bibinfo{title}{{Large-scale cosmic microwave background anisotropies
  and dark energy}}.
\newblock \bibinfo{journal}{\mnras} \bibinfo{volume}{346},
  \bibinfo{pages}{987--993}.
\newblock \DOIprefix\doi{10.1111/j.1365-2966.2003.07144.x},
  \href{http://arxiv.org/abs/astro-ph/0307104}{\tt arXiv:astro-ph/0307104}.
\bibitem[{{Xu}(2014)}]{Xu2014}
\bibinfo{author}{{Xu}, L.}, \bibinfo{year}{2014}.
\newblock \bibinfo{title}{{A New Unified Dark Fluid Model and Its Cosmic
  Constraint}}.
\newblock \bibinfo{journal}{International Journal of Theoretical Physics}
  \bibinfo{volume}{53}, \bibinfo{pages}{4025--4034}.
\newblock \DOIprefix\doi{10.1007/s10773-014-2153-2},
  \href{http://arxiv.org/abs/1210.5327}{\tt arXiv:1210.5327}.
\bibitem[{Xu et~al.(2012)Xu, Lu and Wang}]{xu2012revisiting}
\bibinfo{author}{Xu, L.}, \bibinfo{author}{Lu, J.}, \bibinfo{author}{Wang, Y.},
  \bibinfo{year}{2012}.
\newblock \bibinfo{title}{Revisiting generalized chaplygin gas as a unified
  dark matter and dark energy model}.
\newblock \bibinfo{journal}{The European Physical Journal C}
  \bibinfo{volume}{72}, \bibinfo{pages}{1--6}.
\bibitem[{Zhang et~al.(2006)Zhang, Wu and Zhang}]{zhang2006new}
\bibinfo{author}{Zhang, X.}, \bibinfo{author}{Wu, F.Q.},
  \bibinfo{author}{Zhang, J.}, \bibinfo{year}{2006}.
\newblock \bibinfo{title}{New generalized chaplygin gas as a scheme for
  unification of dark energy and dark matter}.
\newblock \bibinfo{journal}{Journal of Cosmology and Astroparticle Physics}
  \bibinfo{volume}{2006}, \bibinfo{pages}{003}.
\bibitem[{{Zlatev} et~al.(1999){Zlatev}, {Wang} and {Steinhardt}}]{Zlatev1999}
\bibinfo{author}{{Zlatev}, I.}, \bibinfo{author}{{Wang}, L.},
  \bibinfo{author}{{Steinhardt}, P.J.}, \bibinfo{year}{1999}.
\newblock \bibinfo{title}{{Quintessence, Cosmic Coincidence, and the
  Cosmological Constant}}.
\newblock \bibinfo{journal}{\prl} \bibinfo{volume}{82},
  \bibinfo{pages}{896--899}.
\newblock \DOIprefix\doi{10.1103/PhysRevLett.82.896},
  \href{http://arxiv.org/abs/astro-ph/9807002}{\tt arXiv:astro-ph/9807002}.

\end{thebibliography}


\end{document}